\documentclass[twocolumn]{aastex62}

\usepackage{amsmath}
\usepackage{lineno}

\newcommand\Cholla{\emph{Cholla}~}

\newcommand\Lya{Lyman-$\alpha$}

\newcommand\Sim{CHIPS~}

\def\HI{\hbox{\rm H~$\scriptstyle\rm I$}}

\def\HeI{\hbox{He~$\scriptstyle\rm I$}}
\def\HeII{\hbox{He~$\scriptstyle\rm II$}}

\def\hide#1{}
\newcommand\Fmean{$\langle F \rangle$~}
\newcommand\taueff{$\tau_{\mathrm{eff}}$~}

\newcommand\GammaHI{$\Gamma_{\mathrm{HI}}$~}
\newcommand\xHI{$x_{\mathrm{HI}}$~}
\newcommand\xHeII{$x_{\mathrm{HeII}}$~}
\newcommand\taueffH{$\tau_{\mathrm{eff,H}}$~}
\newcommand\taueffHn{$\tau_{\mathrm{eff,H}}$}
\newcommand\taueffHe{$\tau_{\mathrm{eff,HeII}}$~}
\newcommand\taueffHen{$\tau_{\mathrm{eff,HeII}}$}

\graphicspath{{./}{figures/}}

\shorttitle{The Thermal History of the IGM from the \HI\ and \HeII\  \Lya\ Forest}
\shortauthors{Villasenor et al.}

\begin{document}

\title{Inferring the Thermal History of the Intergalactic Medium from the 
Properties of the Hydrogen and Helium \Lya\ Forest}

\correspondingauthor{Bruno Villasenor.}
\email{brvillas@ucsc.edu}

\author[0000-0002-7460-8129]{Bruno Villasenor}
\affiliation{Department of Astronomy and Astrophysics, University of California,
             Santa Cruz, 1156 High Street, Santa Cruz, CA 95064 USA}

\author[0000-0002-4271-0364]{Brant Robertson}
\affiliation{Department of Astronomy and Astrophysics, University of California,
             Santa Cruz, 1156 High Street, Santa Cruz, CA 95064 USA}

\author[0000-0002-6336-3293]{Piero Madau}
\affiliation{Department of Astronomy and Astrophysics, University of California,
             Santa Cruz, 1156 High Street, Santa Cruz, CA 95064 USA}

\author[0000-0001-9735-7484]{Evan Schneider}
\affiliation{Department of Physics and Astronomy \& Pittsburgh Particle Physics, Astrophysics, and Cosmology Center (PITT PACC), University of Pittsburgh, Pittsburgh, PA 15260, USA}

\begin{abstract}
The filamentary network of intergalactic medium (IGM) gas that gives origin to the  \Lya\ forest in the spectra of distant quasars encodes information 
on the physics of structure formation
and the early thermodynamics of 
diffuse baryonic material. Here, we use a massive suite of more than 400 high-resolution cosmological hydrodynamical simulations run with the Graphics Processing Unit-accelerated code {\it Cholla} to study the IGM at 
high spatial resolution maintained over the entire computational volume.
The simulations capture a wide range of possible IGM thermal histories
by varying the photoheating and photoionizing background
produced by star-forming galaxies and active galactic nuclei.
A statistical comparison of synthetic spectra with the observed 1D flux power spectra of hydrogen at redshifts  $2.2\leq z\leq 5.0$
and with the helium \Lya\ opacity at redshifts $2.4<z<2.9$ tightly constrains the photoionization and photoheating history of the IGM.
By leveraging the constraining power of the available \Lya{} forest data to break model
degeneracies, we find
that the IGM experienced two main reheating events over 1.2 Gyr of 
cosmic time.
For our best-fit model, hydrogen reionization completes by $z_\mathrm{R} \approx 6.0$
with a first IGM temperature peak $T_0 \simeq 1.3 \times 10^4 \, \mathrm{K}$, and is 
followed by the reionization of \HeII\ that completes 
by $z_\mathrm{R} \approx 3.0$ and 
yields a second temperature peak of $T_0 \simeq 1.4 \times 10^4 \, \mathrm{K}$.
We discuss how our results can be used  to obtain information on the timing and the sources of hydrogen and helium reionization.
\end{abstract}


\section{Introduction} \label{sec:intro}

The neutral hydrogen and singly-ionized helium components of gas near the cosmic mean density trace the distribution of matter in 
between galaxies and produce a ``forest" of detectable \Lya\ absorption features in the spectra of distant quasars
\citep[e.g.,][]{hernquist1996a,Croft+1998,meiksin09,Slosar+2011,mcquinn2016a,Worseck+2019}. 
The depth, shape, and location of absorption lines in the \Lya\ forest depend on the ionization degree and thermal state of this intergalactic 
medium (IGM), which are controlled by the uncertain UV radiation background produced by star-forming galaxies and active galactic nuclei (AGNs) \citep[e.g.,][]{haardt2012a,Robertson+2015,madau15,faucher-giguere2020a}, 
and on its density and peculiar velocity fields
shaped by gravity \citep{cen1994a}. 
Dark matter provides the backbone of large-scale 
structure in the Universe, a web-like pattern 
present in embryonic form in the overdensity motif of the initial fluctuation field
and 
sharpened by non-linear gravitational dynamics \citep{bond96}. 
The \Lya\ forest traces this underlying ``cosmic web" on scales and at redshifts that cannot be probed by any other observable.
Because of its long cooling time, low-density gas at $z\sim$ 2--5 
that traces the underlying matter distribution 
retains some memory of when and how it was reheated and reionized at 
$z\gtrsim 6$ \citep{Miralda+94}. 
The physics 
that governs the properties of the IGM throughout these epochs
remain similar, as
the evolving cosmic UV emissivity and the transfer of that radiation through 
a medium made clumpy by gravity determine both the details of the reionization process and the thermodynamics of the forest.

Understanding how reionization occurred, the nature of the early sources that drove it, the thermal history and fine-grained properties of hydrogen 
gas in the cosmic web, and how to extract crucial information on the cosmological model from observations of \Lya\ absorption are among the most 
important open questions in cosmology and  key science drivers for numerous major new instruments and facilities. The promise of the \Lya\ forest 
for constraining cosmological physics including  the nature of dark matter and dark energy has 
motivated in part the construction of the Dark Energy 
Spectroscopic Instrument \citep{desi2016a}, 
which measures
absorption line spectra backlit by nearly a million quasars at $z>2$,
and the WEAVE survey \citep{Pieri+2016} which will observe more than 400,000 high-redshift quasars at $z>2$.    
Interpreting such 
observations requires detailed cosmological hydrodynamical simulations that cover an extensive range of uncertain photoionization and photoheating 
histories and consistently maintain high resolution throughout a statistically representative sub-volume of the Universe. 

This paper extends research efforts
directly focused on advancing the state-of-the-art in modeling the IGM physical structure in cosmological simulations while still 
achieving high computational efficiency, thereby providing higher fidelity physical models for interpreting \Lya\ forest data. 
In \citet{villasenor2021a} 
we 
introduced
the Cholla IGM Photoheating Simulations (CHIPS) to investigate how different photoheating histories and cosmological parameters impact the 
structure of the forest. Here, we use a massive suite of more than 400 CHIPS simulations to study the IGM at a resolution of 49$h^{-1}$ ckpc maintained 
over ($50h^{-1}$ cMpc)$^3$ volumes. 
Performed
with the GPU-native MPI-parallelized code {\it Cholla} \citep{schneider2015a}, these simulations span different 
amplitudes and peak redshifts of the \HI\ and \HeII\ photoionization and photoheating rates.

To anticipate the results of our likelihood analysis
constrained by the 1D flux power spectra $P(k)$ measured in eBOSS, Keck, and VLT 
data and the observed \HeII\ \Lya\ forest, we find that scenarios where hydrogen in the cosmic web was fully reionized by star-forming galaxies 
by redshift $z_{\rm R}\approx 6.0$
and the double reionization of helium was completed by quasar sources about 1.2 billion years later are strongly favored by the data. 
Models that reionize hydrogen or helium at earlier or later cosmic times produce too much or too little cold gas, and appear to be inconsistent with 
the observed $P(k)$ and \HeII\ \Lya\ opacity. Our approach differs from previous work in this field in the following aspects:

\begin{enumerate}

\item The simulation grid captures a wide range of possible thermal histories via a four-parameter 
scaling of the amplitude and timing
of the (spatially uniform) 
metagalactic UVB
responsible for determining the ionization states and temperatures of the IGM
(cf. \citealt{Nasir+2016,onorbe2017a}).
We use the physically-motivated model of \citet{puchwein2019a} as a template, and vary
the strength and redshift-timing of their ionization and heating rates.

\item We do not modify, in post-processing, the mean transmitted flux $\langle F\rangle$ in the forest by recalibrating the \Lya\ optical 
depth, nor do we assume or rescale an instantaneous gas temperature-density relation (cf. \citealt{viel2013a,Irsic+2017b,boera2019a,walther2021}).
Indeed, we find from our simulations that the often assumed perfect power-law relationship between the temperature and density of the IGM does not provide
a good approximation over the relevant density and redshift intervals.

\item Our likelihood analysis evaluates
the performance of a given model in matching the observations over the
complete self-consistently evolved reionization and thermal history of the IGM, i.e. over the full redshift
range  $2.2\leq z\leq 5.0$
for the observed 1D flux power spectrum  of hydrogen and over the 
redshift range $2.4<z<2.9$ for the \Lya\ opacity of \HeII. Since the 
properties of the gas at one redshift cannot be disentangled from its properties at previous epochs and the
thermal and ionization structure of the IGM evolve with cosmic time along continuous trajectories, 
the marginalization over the parameter posterior distributions 
should not be
performed independently at each redshift (cf. \citealt{bolton2014a, Nasir+2016, hiss2018a,  boera2019a,   walther2019a, gaikwad2020b}).

\end{enumerate}

This paper aims to find the optimal photoionization and photoheating rates 
that reproduce the observed properties of the hydrogen and helium \Lya\  forest.
In Section \ref{sec:methods} we describe 
the simulations used for this work,
how we apply transformations to the UV background (UVB)
model from \citealt{puchwein2019a} to generate our range of
photoionization and photoheating rates,
and the impact of the different UVB models on the statistics of the forest and the properties of the IGM. We follow by 
presenting the observational data and the methodology for the Bayesian Markov Chain Monte Carlo (MCMC) inference used to constrain the model. Section \ref{sec:discussion}
presents our result for the best-fit model and the comparison of the resulting properties of the forest and the thermal evolution of the IGM to the 
observational determinations and previous inferences. We summarize our results and conclusions in \S \ref{sec:summary}.
In Appendix
\ref{sec:resolution} we discuss resolution effects on the \Lya\ power spectrum $P(k)$ from our simulations.
A quantitative study of the
impact on $P(k)$ from rescaling the effective optical of the skewer sample 
is presented 
in Appendix \ref{sec:rescale_tau}. In Appendix \ref{sec:covariance_matrices} we show the variation in the
covariance matrix of the \Lya\ power spectrum from our simulations.    
We discuss in Appendix \ref{sec:colder_IGM} how possible alterations to our
model can modify the predicted temperature history of the IGM.
Finally,
Appendix \ref{sec:phase_diagram} analyzes
the accuracy of assuming a power-law relation for the density-temperature distribution of the gas in our simulations. 

\section{Methodology} \label{sec:methods}

For the study presented here, we compare the observed statistics of the \Lya\ forest to simulations that apply different models for the metagalactic UVB. 
In this section we briefly describe our simulation code and the method to 
extract \Lya\ spectra from the simulations. We then describe 
our simulation grid and the effects that the different UVB models have on the properties of the IGM. Finally we present the observational measurements 
and the inference method used to constrain our model for the UVB
photoionization and photoheating rates. 

\subsection{Simulations} \label{sec:simulation}

The simulations used for this work were run with the 
cosmological hydrodynamics code \Cholla \citep{schneider2015a, villasenor2021a}. 
\Cholla evolves the equations of hydrodynamics on a uniform Cartesian grid using a finite volume approach with 
a second-order Godunov scheme \citep{Colella+1984}. The simulations track the ionization states of hydrogen and helium 
given by the photoionization 
from the UVB, recombination with free electrons and collisional ionization. The non-equilibrium  H+He chemical network is evolved
simultaneously with the hydrodynamics using the GRACKLE library \citep{smith2017a}. We assume a spatially uniform, time-dependent UVB  
in the form of redshift-dependent photoionization rates per ion $\Gamma$ and photoheating rates per ion $\cal{H}$ for neutral hydrogen \HI, neutral 
helium \HeI,  and singly ionized helium \HeII. For a detailed description of the simulation code we refer the reader to the methodology 
section presented in \cite{villasenor2021a}.   

The initial conditions for our simulations were generated using the MUSIC code \citep{Hahn+2011_Music} for a flat $\Lambda$CDM cosmology with parameters $H_0$ = 67.66 km s$^{-1}$, $\Omega_m$ = 0.3111, $\Omega_{\Lambda}$ = 0.6889, $\Omega_b$ = 0.0497, $\sigma_8$ = 0.8102, 
and $n_s$ = 0.9665, consistent with the constraints from \cite{Planck_collaboration_2020}. In future work, we plan to extend our analysis
and include variation of the cosmological parameters \citep{Bird+2019, Ho+2021}.   
Unless otherwise stated the volume and numerical size of 
our simulations correspond to $L$ = 50 $h^{-1}$Mpc and $N$ = 2$\times$1024$^3$ cells and particles. The initial conditions for all runs were 
generated from identical random number seeds to preserve the same amplitude and phase for all initial Fourier modes across the simulation suite.

\begin{figure*}
\includegraphics[width=\textwidth]{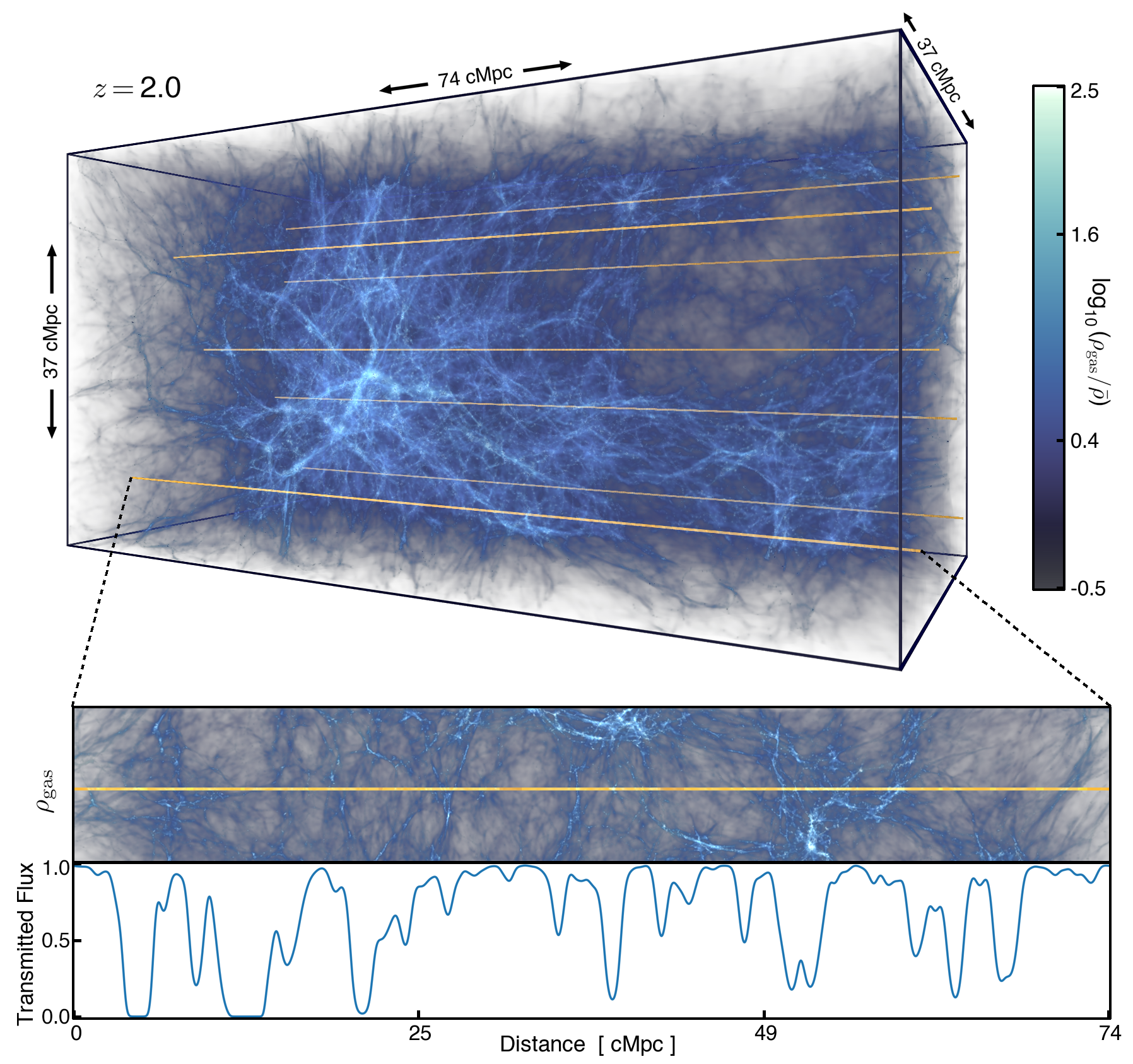}
\caption{Large-scale distribution of gas density (top) from one of our highest-resolution cosmological simulations 
($L$=50 $h^{-1}$Mpc, $N$=2$\times$2048$^3$ cells and particles) at redshift $z$=2 and a set of skewers crossing the simulated box (yellow lines). 
The bottom panels show the density of the gas surrounding a selected line of sight and the \Lya\ transmitted flux along the skewer. Absorption lines in the forest probe the \HI\ column density, the peculiar velocity, and the temperature of the gas along the line of sight.}    
\label{fig:density_3D_skewers}
\end{figure*}

\subsection{Synthetic \Lya\ Spectra }

The \Lya\ forest sensitively probes
the state of the baryons in the IGM, 
and
absorption lines from the forest reflect the \HI\ content and the temperature 
of the gas in the medium. To compare the properties of the IGM in our simulations directly to observations, we extract synthetic hydrogen \Lya\ forest spectra 
from the simulated boxes by measuring the \HI\ density, temperature, and peculiar velocity of the gas along 12,228 skewers through the simulation volume, 
using 4096 skewers along each axis of the box. The optical depth $\tau$ as a function of velocity $u$ along each skewer is computed by integrating the product 
of the \Lya\ scattering cross section and the number density of neutral hydrogen along the line of sight as described in \cite{villasenor2021a}.

The transmitted flux $F$ is computed from the optical depth $\tau$ along the skewers according to $F=\exp(-\tau)$. The power spectrum of the 
transmitted flux $P(k)$ is calculated as the average amplitude of the one-dimensional Fourier transform of the flux fluctuations $\delta_{F}(u)$, 

\begin{equation}
\delta_{F}(u) \equiv \frac{ F(u)- \langle F \rangle}{ \langle F \rangle }  ,
\label{eq:delta_F}
\end{equation}
\noindent
where \Fmean is the average transmitted flux over the skewer sample at a given redshift (see \S5.4 from \citealt{villasenor2021a}  for a detailed description). Similarly, we extract the flux $F_{\mathrm{HeII}}$
transmitted through the \HeII\ \Lya\ forest from the simulations, and compute the \HeII\ effective optical depth 
as \taueffHe$= - \ln \, \langle F_{\mathrm{HeII}} \rangle $.

Figure \ref{fig:density_3D_skewers} (top) shows the gas density distribution at redshift $z=2$ from a section taken from one of our highest-resolution 
($L$=50 $h^{-1}$Mpc, $N=2\times$2048$^3$ cells and particles) simulations, where several skewers crossing the simulated box are shown as yellow lines. 
The bottom panels show the gas density surrounding a selected line of sight and the transmitted hydrogen \Lya\ flux along the skewer. The absorption lines in the forest probe the \HI\ column density, the peculiar velocity, and the temperature of the gas along the line of sight.

\begin{figure*}
\includegraphics[width=\textwidth]{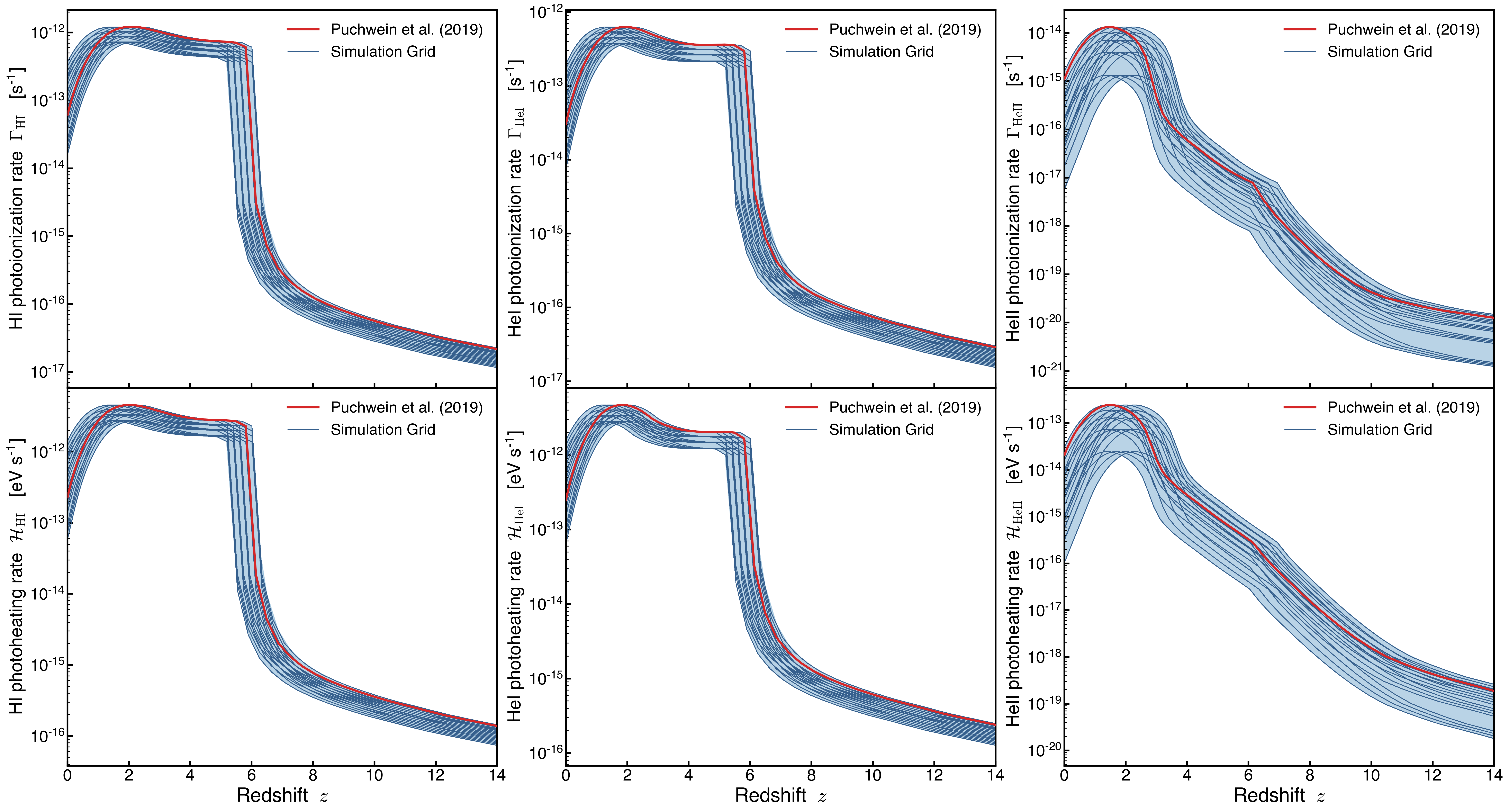}
\caption{Photoionization ($\Gamma$, top) and photoheating ($\cal{H}$, bottom) 
rates for neutral hydrogen (\HI, left), neutral
helium (\HeI, center) and singly ionized helium (\HeII, right) from the reference P19 model \citep{puchwein2019a} (red line) along with the 
photoionization and photoheating rates (blue lines) used for the 400 simulations of the CHIPS grid. The modified rates are generated by rescaling 
and shifting the reference P19 model as described in \S \ref{sec:sim_grid}.}    \label{fig:uvb_rates}
\end{figure*}

\subsection{Photoionization and Photoheating Rates} \label{sec:uvb_rates}

The ionization and thermal evolution of the IGM is primarily determined by the radiation emitted by star-forming galaxies and AGNs over cosmic history 
\citep{mcquinn2016a, Upton-Sanderbeck+2016, onorbe2017a}. The photoionization and photoheating rates adopted in our simulations are computed from the intensity 
of the background radiation field, which is in turn determined by the emissivity of the radiating sources and the opacity of the IGM to ionizing photons. 
Recent models of the UVB \citep{puchwein2019a, khaire2019a, faucher-giguere2020a}, when applied to cosmological simulations, result in a hydrogen 
reionization era that completes by $z \sim$ 6--8 in agreement with observational constraints \citep{davies2018b, Planck_collaboration_2020}. 

The updated model for the photoheating and photoionizing background presented in \citet[][hereafter P19]{puchwein2019a} adopts an improved 
treatment of the IGM opacity to ionizing radiation that consistently captures the transition from a neutral to an ionized IGM. To compute the intensity of the 
background radiation, the P19 model employs recent determinations of the ionizing emissivity due to stars and AGNs and of the \HI\ absorber column density 
distribution, and assumes an evolving escape fraction of ionizing radiation from galaxies into the IGM that reaches 18\%. When the P19 model is applied 
in cosmological simulations, hydrogen reionization completes at $z \sim 6$ consistently with recent measurements 
\citep{Becker+2001, Bosman_2018, Becker+2021, Qin+2021}. However, the subsequent evolution of the \Lya\ forest 
spectra measured in simulations 
that use the P19 model fail to reproduce the observed properties of the forest \citep{villasenor2021a} and, in particular, do not agree with the observed power spectrum of the \Lya\ transmitted flux over the redshift range $2.2 \le z \le 5.0$. This work aims to present a new model photoionization 
and photoheating rates that result in an evolution of the IGM consistent with the observational measurements of the \Lya\ flux power spectrum and the 
\HeII\ effective optical depth.

\subsection{Simulation Grid}\label{sec:sim_grid}

To determine ionization and heating histories that result in properties of the IGM consistent with the observed \Lya\ flux power spectrum and \HeII\ effective opacity, we perform an unprecedented grid consisting of 400 cosmological simulations as a direct extension of the Cholla IGM Photoheating 
Simulations (CHIPS) suite originally presented in \cite{villasenor2021a}. Each simulation in the CHIPS grid applies different photoionization and photoheating 
rates to model a variety of reionization and thermal histories, and thereby produce different statistical properties for the \Lya\ forest. 
To generate different representations of the UVB, we modify the reference model from \cite{puchwein2019a} by rescaling the photoionization  and photoheating  
rates ($\Gamma$ and $\cal{H}$ respectively) by a constant factor $\beta$ and shifting the redshift dependence of the rates by an offset $\Delta z$. The two 
transformations are expressed as

\begin{equation}
\begin{aligned}
\Gamma(z) &\rightarrow \beta \, \Gamma^{\mathrm{P19}}\,( z  - \Delta z), \\ 
\mathcal{H}(z) & \rightarrow \beta \, \mathcal{H}^{\mathrm{P19}}( z - \Delta z).
\end{aligned}
\label{eq:rates_transformation}
\end{equation}
Since the photoionization and photoheating rates for both \HI\ and \HeI\ are dominated  by the same sources, namely star-forming galaxies at $z\gtrsim 5$ and AGNs at lower redshifts, and the radiation that ionizes both species is absorbed by intergalactic hydrogen, we modify the \HI\ and \HeI\ photoionization and photoheating rates jointly by applying the transformations described by Eqs. (\ref{eq:rates_transformation}), 
scaling and shifting by the parameters $\beta_{\mathrm{H}}$ and $\Delta z_{\mathrm{H}}$ respectively. 
\HeII\ is reionized later in cosmic history primarily by 
the extreme UV radiation emitted by AGNs, and
we rescale and redshift-offset the photoionization and photoheating rates associated with \HeII\ by a
second set of parameters 
$\beta_{\mathrm{He}}$ and $\Delta z_{\mathrm{He}}$.  Hence, each modified 
UVB model is characterized by the parameter vector 
$\theta= \{ \beta_{\mathrm{H}}, \,\Delta z_{\mathrm{H}}, \, \beta_{\mathrm{He}}, \,\Delta z_{\mathrm{He}} \} $.
The different photoionization and photoheating histories span
all the combinations of the parameter values presented in Table \ref{tab:sims_params}.

The rescaling parameters $\beta_{\mathrm{H}}$ and  $\beta_{\mathrm{He}}$ control the intensity of the background radiation, determine the efficiency 
with which \HI\ and \HeII\ become ionized, and govern energy input into the IGM in the form of photoheating during the epochs of non-equilibrium reionization for hydrogen and helium. After reionization completes and the gas 
reaches photoionization equilibrium, 
the balance between ionizations from the background radiation 
and 
recombinations with free electrons
determines the ionization state of \HI\ and \HeII.
At equilibrium, the ionized fraction of \HI\ and \HeII\ is proportional to the photoionization rates $\Gamma_{\mathrm{HI}}$ and 
$\Gamma_{\mathrm{HeII}}$ respectively, and inversely proportional to the temperature-dependent radiative recombination rates $\alpha_{\mathrm{HII}}(T)$ and 
$\alpha_{\mathrm{HeIII}}(T)$. Therefore, by rescaling the photoionization and photoheating rates, we modify the evolution of the temperature and the ionization state of the 
gas in the IGM during and after \HI\ and \HeII\ reionization.            

The parameters $\Delta z_{\mathrm{H}}$ and $\Delta z_{\mathrm{He}}$ shift the redshift dependence of the photoionization and photoheating rates by a constant offset, affecting
the timing of \HI\ and \HeII\ reionization. In general, an offset of $\Delta z_{\mathrm{H}} > 0$ or $\Delta z_{\mathrm{He}} > 0$  moves \HI\ or \HeII\ reionization to higher redshift
and earlier cosmic time relative to
the reference P19 model.
Negative values of $\Delta z_{\mathrm{H}}$ or 
$\Delta z_{\mathrm{He}}$ shift reionization to lower redshift and later cosmic times.
The offset in redshift of the models also 
affect the properties of the IGM after \HI\ and \HeII\ reionization 
complete,
as the photoheating and
photoionization rates at a given redshift are generally modified 
when 
$\Delta z_{\mathrm{H}}\ne0$ or 
$\Delta z_{\mathrm{He}}\ne0$.

\begin{deluxetable}{cc}[h]
\tablenum{1}
\caption{\Sim Simulation Grid\label{tab:sims_params}}
\tablewidth{\columnwidth}
\tablehead{
	\colhead{Parameter} & \colhead{Parameter Values} \\[-16pt]
}
\startdata
$\beta_{\mathrm{H}}$ & 0.60, 0.73, 0.86, 1.00 \\
$\Delta z_{\mathrm{H}}$ & -0.6, -0.4, -0.2, 0.0, 0.2 \\
$\beta_{\mathrm{He}}$ & 0.10, 0.30, 0.53, 0.76, 1.00 \\
$\Delta z_{\mathrm{He}}$ & -0.1, 0.2, 0.5, 0.8 \\
\enddata
\tablecomments{The parameters $\beta_\mathrm{H}$ and $\Delta z_\mathrm{H}$ determine the amplitude and redshift offset 
of the \HI\ and \HeI\ photoionization and photoheating rates, while $\beta_\mathrm{He}$ and $\Delta z_\mathrm{He}$  rescale and offset the \HeII\ rates.}
\end{deluxetable}
Figure \ref{fig:uvb_rates} 
shows the photoionization  and photoheating rates from the reference model by \cite{puchwein2019a} together with the modified rates adopted in the 400 simulations of the CHIPS grid.
In \cite{villasenor2021a}, we presented a comparison of the statistical properties of the \Lya\ forest and the thermal history of the IGM that result from a high-resolution 
simulation 
using the UVB model from \cite{puchwein2019a}.
We concluded that, in general, the gas in the simulation was 
too highly ionized after hydrogen reionization and possibly too hot during the epoch of helium reionization to be compatible with the observed statistics of the forest 
and other inferences of the thermal state of the IGM.
We therefore do not include values of $\beta_{\mathrm{H}}>1$ or $\beta_{\mathrm{He}}>1$ 
in our grid, as such models would result in 
overall
higher ionization fractions and 
temperatures of the IGM compared with the P19 case.  

The simulations were run on the Summit system (Oak Ridge Leadership Computing Facility at the Oak Ridge National Laboratory). 
Each simulation was performed
on 128 GPUs and completed in less than  
two wall clock hours.
The cost of the entire grid of computations was only $\sim$16,000 node hours.  
This work demonstrates that by taking advantage of an efficient code like \Cholla and a capable system like Summit, future studies of the IGM using thousands of cosmological simulations are now possible.

\begin{figure*}
\includegraphics[width=\textwidth]{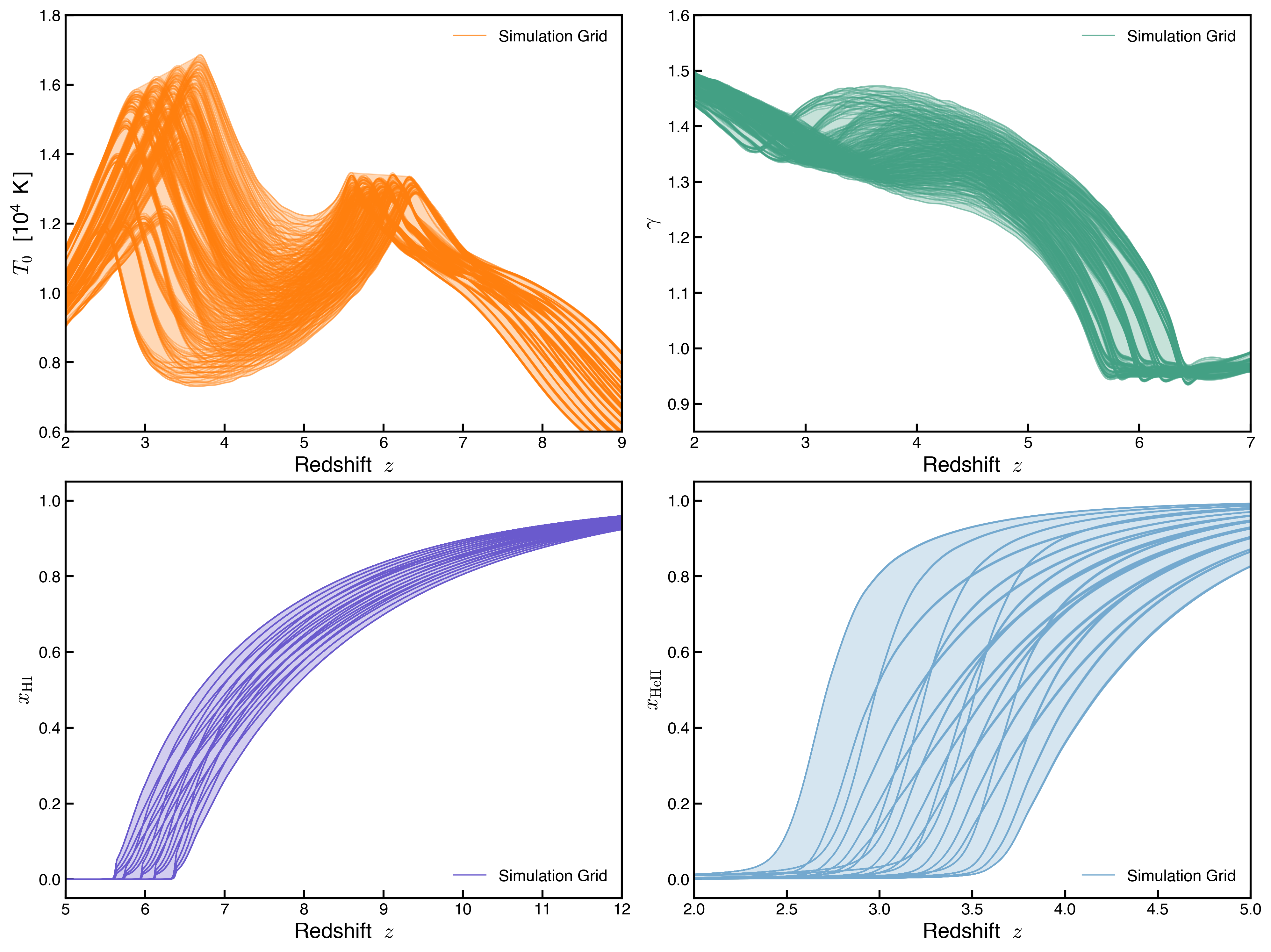}
\caption{
Evolution of global properties of the IGM computed form the 400 CHIPS simulations. The simulations evolve under different 
photoionization and photoheating rates resulting in a large variety of ionization and thermal histories of the IGM. The top panels show the 
the temperature, $T_0$, of intergalactic gas at the mean density (left) and the index $\gamma$ from the power-law density-temperature relation    
$T(\Delta) = T_0 \Delta ^{\gamma - 1}$ (right). The bottom panels show the volume-weighted average of the neutral hydrogen fraction \xHI (left) and the singly 
ionized helium fraction \xHeII (right). The amplitude and timing of the rates impact the 
thermal state of the IGM during \HI\ and \HeII\ reionization. Simulations with higher values of $\beta_\mathrm{He}$ result in a higher temperature peak during \HeII\ reionization 
($2.5 \lesssim z \lesssim 3.8 $) and for simulations with $\Delta z_\mathrm{He} >0$ the epoch of \HeII\ reionization is shifted to earlier epochs. Analogously, 
negative values of $\Delta z_\mathrm{H}$ move the timing of \HI\ reionization to later epochs and simulations with different $\beta_\mathrm{H}$ show a 
different temperature peak during \HI\ reionization at $z\sim 5.6 - 6.3$.}
\label{fig:grid_IGM_prop}
\end{figure*}

\begin{figure*}
\includegraphics[width=\textwidth]{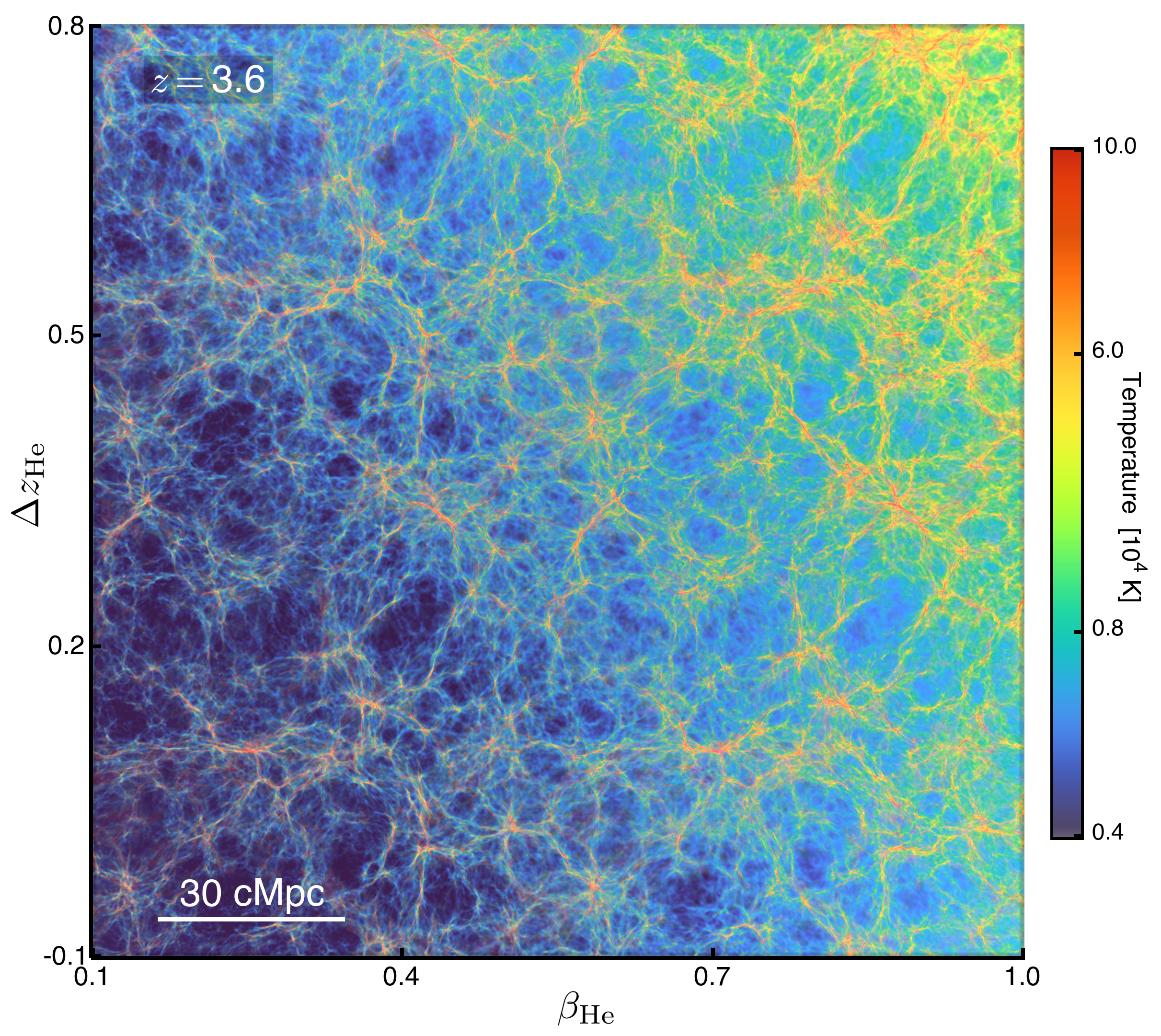}
\caption{Gas temperature from a slice through the IGM at $z=3.6$ in a subset of 20 simulations with different \HeII\ reionization scenarios. An increase 
in the parameters $\beta_{\mathrm{He}}$ and $\Delta z_{\mathrm{He}}$ corresponds to higher \HeII\ photoheating and a shift of the \HeII\ reionization epoch to 
earlier cosmic times (closer to $z\sim 3.6$) respectively. Either effect increases the temperature of the IGM at $z\sim3.6$.}
\label{fig:temperature_slice}
\end{figure*}

\subsection{Effects of UVB Models on the IGM Properties}

The different photoionization and photoheating histories adopted in our 
simulations affect the ionization state of hydrogen and helium and  the temperature
of the IGM. Figure \ref{fig:grid_IGM_prop} shows the redshift evolution of the
global properties of the IGM for each of the simulated histories. The top panels show
the temperature of gas at mean density $T_0$ (left) and the index $\gamma$ (right) of the power-law density-temperature 
relation $T(\Delta) = T_0 \Delta ^{\gamma - 1}$, where 
$\Delta = \rho_\mathrm{gas}/\bar{\rho}$. The bottom panels show the volume weighted average 
fraction of neutral hydrogen \xHI (left) and singly ionized helium \xHeII (right).

As hydrogen becomes ionized at $z\gtrsim 5.5$ the gas in the IGM experiences a monotonic increase of $T_0$
while showing a close to isothermal distribution $\gamma \sim 1$. After hydrogen reionization ends at $z \sim$ 5.5--6.5, 
the gas cools primarily through the adiabatic expansion of the Universe. During this period the low-density gas cools
faster and $\gamma$ increases. This first epoch of cooling ends with the onset of helium (\HeII) reionization 
from the extreme UV radiation emitted by AGNs at $z \lesssim 4-5$ which reheats the IGM, increasing $T_0$ and decreasing $\gamma$.
After the double reionization of helium completes ($z\sim$ 2.5--3.5) the IGM cools monotonically  
by adiabatic expansion increasing $\gamma$ for a second time. Because of these two distinct photoheating epochs, 
the thermal state of the IGM in our simulations is more sensitive
to variations in the hydrogen photoheating/photoionization parameters 
$\beta_{\mathrm{H}}$ and $\Delta z_{\mathrm{H}}$
at $z\gtrsim 5$, and more sensitive to the parameters 
$\beta_{\mathrm{He}}$ and $\Delta z_{\mathrm{He}}$ at
$z\lesssim 5$ during the epoch of helium reionization.

For simulations with $\Delta z_\mathrm{H} <0$ the temperature peak from hydrogen reionization is shifted to later times (lower redshift) 
and the amplitude of the temperature peak depends
on the value of $\beta_\mathrm{H}$. 
Analogously, the parameters $\beta_\mathrm{He} $ and $\Delta z_\mathrm{He}$
determine the amplitude and timing of the second peak in $T_0$ caused by 
helium reionization. Positive values of $\Delta z_\mathrm{He}$ move helium reionization 
to higher redshifts compared with the reference P19 model, and higher values of $\beta_\mathrm{He}$ 
produce a higher peak in $T_0$ during the
epoch $2.5 \lesssim z \lesssim 3.8$.

Variation in the timing of H and He reionization changes the cooling periods during which the power-law index $\gamma$ 
increases.
The different tracks of $\gamma$ in our simulation grid 
then arise primarily from the different values of
$\Delta z_\mathrm{H}$ and $\Delta z_\mathrm{He}$ adopted. In future work we plan to expand the flexibility of our simulations to sample 
the thermal state of the IGM by introducing density dependent photoheating rates which will allow for larger variation in the 
evolution of $\gamma$.

\begin{figure*}
\includegraphics[width=\textwidth]{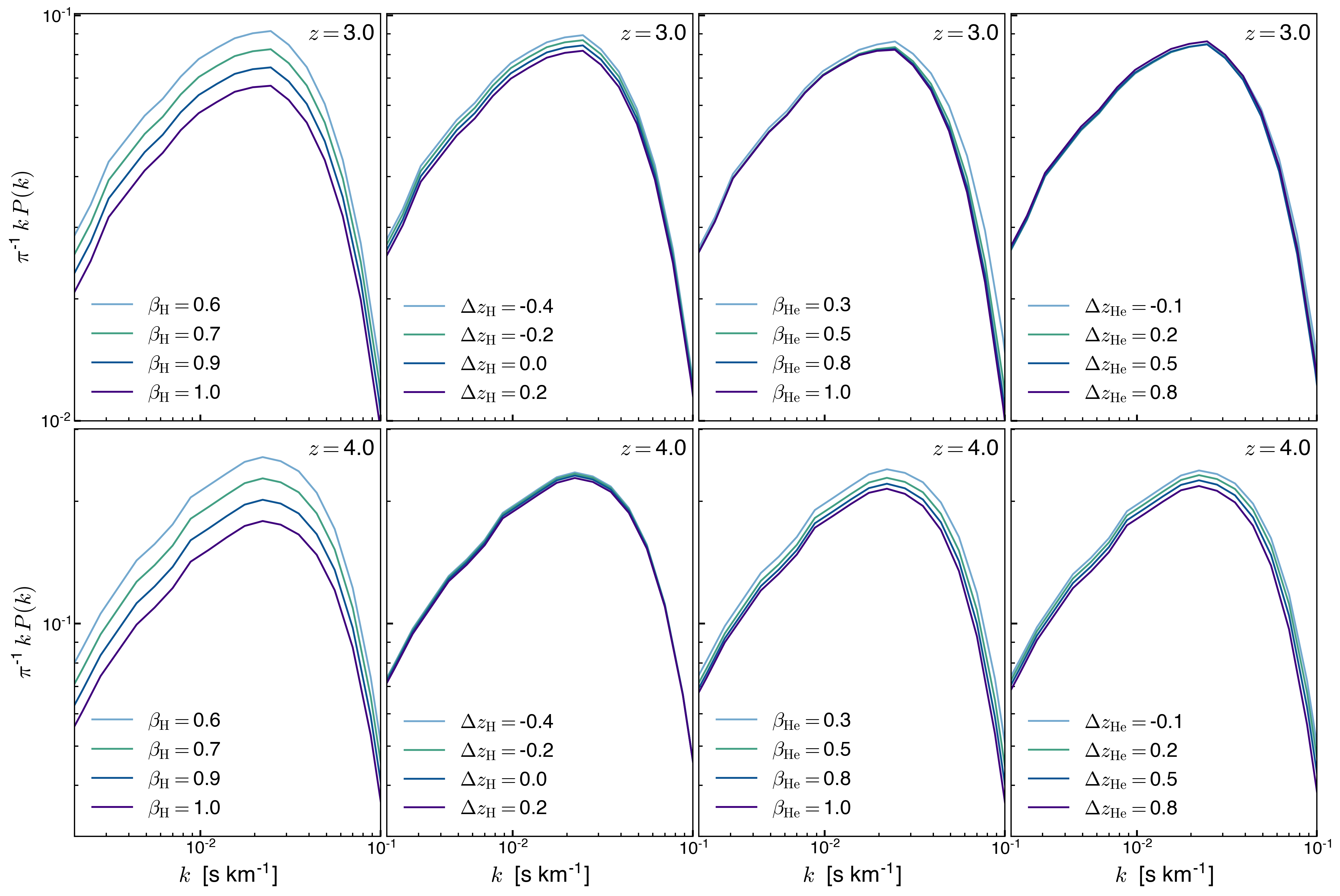}
\caption{Sensitivity of the \Lya\ flux power spectrum $P(k)$ to independent variations 
of the parameters 
$\theta= \{ \beta_{\mathrm{H}}, \,\Delta z_{\mathrm{H}}, \, \beta_{\mathrm{He}}, \,\Delta z_{\mathrm{He}} \} $ for redshifts $z=3$ (top) and $z=4$ (bottom). Independent 
changes of each parameter have different effects on the redshift-dependent $P(k)$. After hydrogen reionization completes,  differences in the power 
spectrum  at $z\lesssim 5.5$ arise from changes in the ionization state and temperature 
of the IGM. Variation of the parameters $\beta_\mathrm{H}$ and $\Delta z_\mathrm{H}$ mostly 
affect the ionization state of hydrogen and therefore the overall normalization of $P(k)$. 
Changes in the parameters $\beta_\mathrm{He}$ and $\Delta z_\mathrm{He}$ 
impact $P(k)$ through their effect on the temperature of the gas during and after helium reionization, as variations in the thermal state of the IGM control the ionization 
fraction of hydrogen by its effect on the recombination rate $\alpha_\mathrm{HII}(T)$, 
and lead to the Doppler broadening of absorption lines and the smoothing of 
density fluctuations that suppress small-scale power ($ k \gtrsim 0.02 \,\, \mathrm{s \, km^{-1}}$).}    
\label{fig:parameters_ps}
\end{figure*}

The effects on the temperature of the IGM from the different helium reionization scenarios in our simulations are illustrated in Figure \ref{fig:temperature_slice}. The image 
displays the gas temperature of a slice through the IGM at $z=3.6$ generated from a subset of 20 simulations that vary the parameters $\beta_{\mathrm{He}}$ and 
$\Delta z_{\mathrm{He}}$ 
controlling the \HeII\ photoionization and photoheating rates. 
Increases in $\beta_{\mathrm{He}}$ and $\Delta z_{\mathrm{He}}$ correspond to a larger 
extreme UV background from AGNs and to a shift of the epoch of helium reionization to earlier cosmic times, respectively. Either effect causes the temperature of the 
IGM to increase at $z \sim 3.6$. 
Decreasing the \HeII\ photoheating rates or shifting helium reionization to later cosmic times (toward $z\sim 2.8$) 
decreases the temperature of IGM gas at $z\sim 3.6$. 

\begin{figure*}
\includegraphics[width=\textwidth]{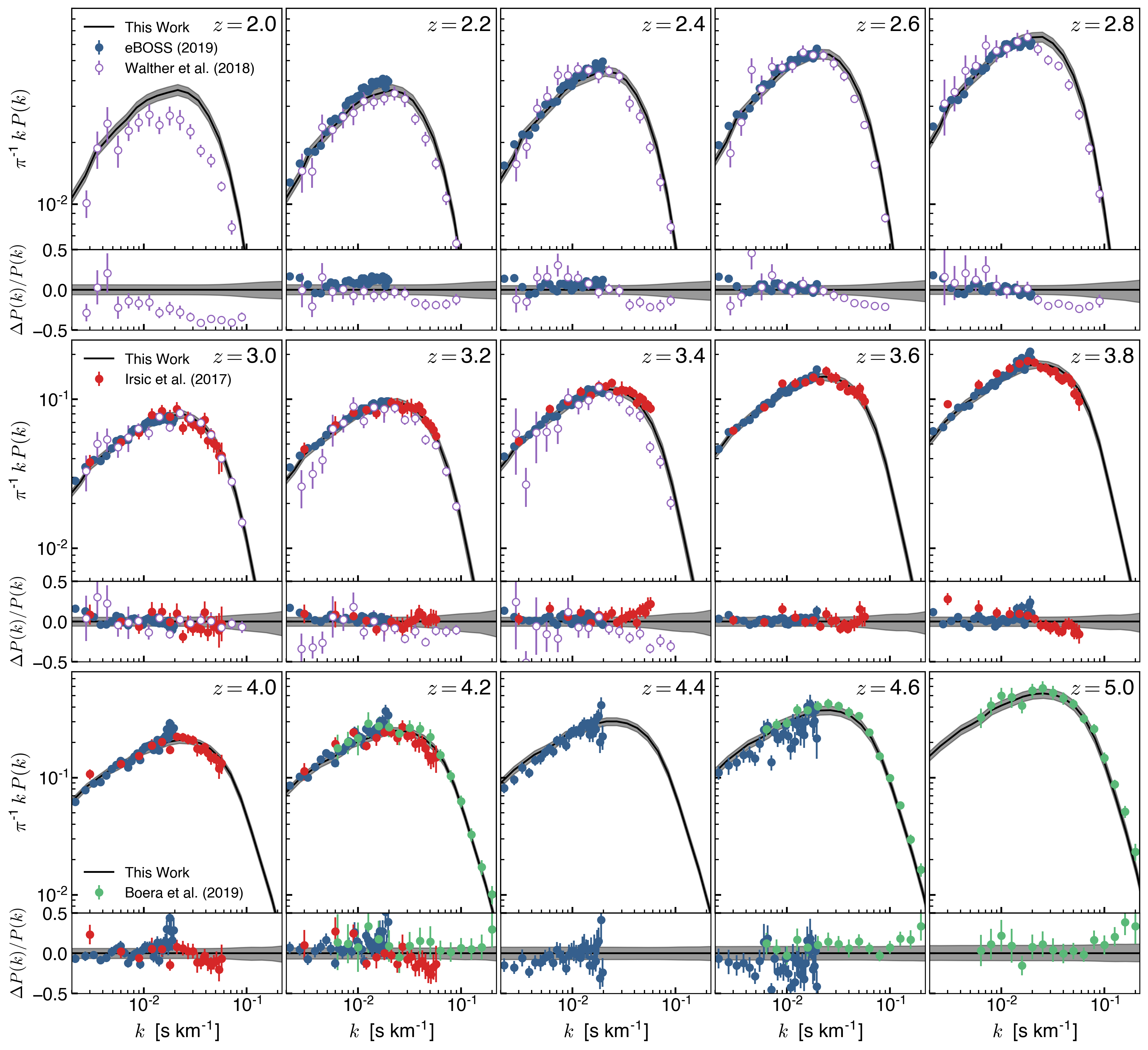}
\caption{The transmitted flux power spectrum $P(k)$ 
from observations by eBOSS \citep{Chabanier+2019}, Keck Observatory and the Very Large Telescope \citep{Irsic+2017a, boera2019a} 
used to constrain models of the cosmic photoionization and photoheating history. 
The best-fit evolution of $P(k)$ marginalized over the posterior distribution of the parameters $\theta= \{ \beta_{\mathrm{H}}, \,\Delta z_{\mathrm{H}}, \, \beta_{\mathrm{He}}, \,\Delta z_{\mathrm{He}}\}$ is shown with black curves, along 
with 95\% confidence intervals (shaded bands). 
The fractional differences from the observations and the best-fit model are shown in the bottom part of each panel.
Overall, the best-fit $P(k)$ is in good agreement with the large-scale power spectrum from eBOSS for $2.4 \leq z \leq 4.2$, 
and with the intermediate scales data from \cite{Irsic+2017a} at $3.0 \leq z \leq 4.2$. 
Our best-fit results also agree with the measurements from \cite{boera2019a} at  $4.2 \leq z \leq 5.0$, showing 10--30\% differences mostly on the smallest scales ($0.1 - 0.2\,\, \mathrm{s\, km^{-1}}$) and suggesting that the temperature of the IGM at this epoch could be slightly overestimated by 
the model. We also show the $P(k)$ determinations by \cite{walther2018a} for comparison.
Owing to discrepancies with the eBOSS results on large scales, we have not included the \cite{walther2018a} data points in our MCMC analysis.}
\label{fig:power_spectrum_all}
\end{figure*}

\begin{figure}
\includegraphics[width=0.47\textwidth]{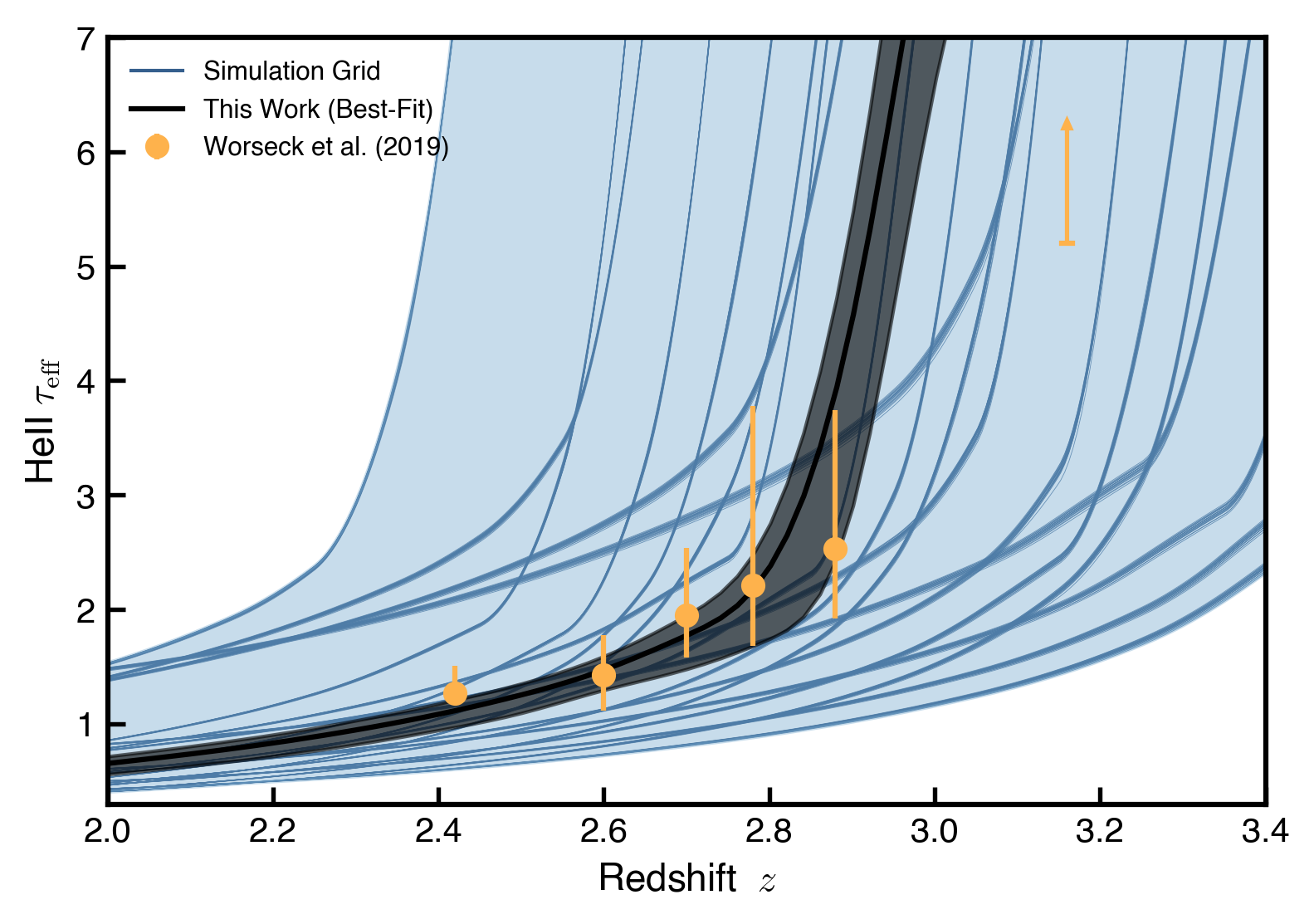}
\caption{Evolution of the singly ionized helium (\HeII) effective optical depth \taueffHe from our simulation grid (blue lines), along with the best-fit model (black line) and the 95\% 
confidence interval (gray area) obtained from our MCMC marginalization.
The orange points show the observational measurements of \taueffHe \citep{Worseck+2019}. While only data in the redshift 
range $2.4 \lesssim z \lesssim 2.9$ were used as constraints for our statistical analysis, the
observed lower limits at $z > 3$ are  consistent with the model results.}
\label{fig:tau_HeII}
\end{figure}

\subsection{Effects of UVB Models on the \Lya\ Forest Power Spectrum}

The statistical properties of the \Lya\ forest provide insight into the state of the baryons in 
the IGM.  The effective optical of the forest \taueffH$= - \ln \, \langle F \rangle$ provides a 
global measurement of the overall \HI\ content of the gas in the IGM,
probes the hydrogen ionization fraction, and allows for estimates of 
the intensity of the ionizing background radiation.
The power spectrum $P(k)$ of the flux transmitted through the forest contains more information encoded across 
different spatial scales. On scales larger than a few 
Mpc the $P(k)$ is sensitive to the ionization fraction of 
hydrogen in the IGM and provides
information similar to $\tau_{\rm eff,H}$.
This connection makes $P(k)$ and \taueffH 
a dependent
pair of measurements, and  
\S\ref{sec:rescale_tau} 
presents a detailed analysis 
about 
the effects that variations in \taueffH induce in $P(k)$.
On scales smaller than a few comoving Mpc,
structure in the forest is suppressed by pressure smoothing of the gas density fluctuations 
as well as Doppler broadening of the absorption lines.
These effects cause a cutoff in the dimensionless power spectrum $\Delta^2(k) = \pi^{-1} k P(k)$
for $k \gtrsim 0.02\,\mathrm{s\,km^{-1}}$,
making the flux power spectrum at intermediate and small scales a sensitive probe of the thermal state of IGM gas.

The different ionization and thermal histories produced by the range of photoionization and photoheating 
rates adopted in our simulations
manifest as variations in the flux power spectrum of the \Lya\ forest. The effects on $P(k)$ from changing each of the four parameters 
$\beta_{\mathrm{H}}$, $\Delta z_{\mathrm{H}}$, $\beta_{\mathrm{He}}$, or $\Delta z_{\mathrm{He}}$ independently is shown in Figure \ref{fig:parameters_ps} 
for redshifts $z=3$ (top) and $z=4$ (bottom). 
The variation in $P(k)$ measured from our simulation grid over the redshift range 
$2 \lesssim z \lesssim 5$ 
can be attributed
mainly to three 
physical effects.
First, since hydrogen is in photoionization equilibrium after \HI\ reionization, 
changes to the photoionization rate \GammaHI
from rescaling by $\beta_\mathrm{H}$ or applying a shift
$\Delta z_\mathrm{H}$ alter the ionization fraction of hydrogen. This
alteration globally affects the hydrogen effective optical depth \taueffH and, as a result, the 
overall normalization of $P(k)$ changes. 
Second, changes in the temperature of the IGM from the different hydrogen and helium reionization scenarios
alter the recombination coefficient $\alpha_\mathrm{HII}(T)$ in the IGM. In turn, changes to the recombination
rate adjust the ionization fraction of hydrogen in the IGM and thereby the normalization of $P(k)$.
Third, the different thermal histories of the IGM 
affect $P(k)$ on small scales through Doppler broadening of the absorption lines and the pressure smoothing of the density fluctuations.
As shown in Figure \ref{fig:parameters_ps}, 
the parameters $\beta_\mathrm{H}$ 
and $\Delta z_\mathrm{H}$ mainly 
influence
the normalization of $P(k)$ by changing the overall ionization fraction in the IGM, while the parameters $\beta_\mathrm{He}$ and 
$\Delta z_\mathrm{He}$ change the temperature of the IGM
and thereby affect both the normalization and small-scale shape of $P(k)$.                          

\subsection{Observational Data}
\label{sec:observational_data}

For comparison with our simulations,
we use the observational determinations of the flux power spectrum measured by the extended 
Baryon Oscillation Spectroscopy Survey \citep[eBOSS;][]{Chabanier+2019} and separate measurements with the Keck Observatory and the Very Large Telescope 
\citep{Irsic+2017a, boera2019a}. 
The power spectrum estimates from \cite{Chabanier+2019} probe mostly large scales 
($ 0.001 \lesssim k \lesssim 0.02 \,\,\mathrm{s\,km^{-1}}$) in the redshift range $2.2 < z < 4.6$. The determinations from \cite{Irsic+2017a} overlap with 
the eBOSS measurements on the large scales,
albeit with lower precision,
and extend to intermediate scales ($ 0.003 \lesssim k \lesssim 0.06 \,\,\mathrm{s\,km^{-1}}$)
for redshifts $3.0 < z < 4.2$. 
The data from \cite{boera2019a} cover intermediate to small scales ($ 0.006 \lesssim k \lesssim 0.2 \,\,\mathrm{s\,km^{-1}}$) 
over the redshift range $4.2 < z < 5.0$. 
The combined data set spans a large redshift range from $z=2.2$ to $z=5.0$ and a wide range of scales, and is shown along with our best-fit model $P(k)$ in Figure \ref{fig:power_spectrum_all}.   

Figure
\ref{fig:power_spectrum_all} also shows
the observational measurements of $P(k)$ presented by \citet[][purple empty points]{walther2018a}
for the redshift range $3.0 \leq z \leq 3.4$.
We find that, in the overlapping range of scales ($0.003 \lesssim k \lesssim 0.02 \,\, \mathrm{s \, km^{-1}}$) and redshift 
($2.2 \lesssim z \lesssim 3.4$), the estimates from \cite{walther2018a} show significant differences with those  from eBOSS \citep{Chabanier+2019}.
The normalization and, in some cases, the shape of the large-scale $P(k)$ appear to be inconsistent between the two datasets. For several redshift bins 
(e.g. $z=2.4$ and $z=3.2$),
a simple renormalization applied to 
the \citet{walther2018a}
power spectrum  would not be sufficient to match the large-scale measurements from eBOSS. Because of this discrepancy, we have not included the \cite{walther2018a} $P(k)$ determinations in  our MCMC analysis, and we 
show them in 
Figure \ref{fig:power_spectrum_all} only for comparison with our modeling and other data sets.     

To obtain a better determination of the \HeII\ photoionization and photoheating rates, we complement the power spectrum comparison with observational 
measurements of the helium effective optical depth \taueffHe \citep{Worseck+2019} over the redshift range $2.4 \lesssim z \lesssim 2.9$  as additional constraints on our model. 
The data are shown in Figure \ref{fig:tau_HeII}
along with the corresponding evolution of \taueffHe from our simulation grid and the best-fit model from our analysis. We do not include the observational lower 
limits at $z>3$ as constraints in our MCMC analysis, but our best-fit model is consistent with those limits.

\subsection{Systematic Uncertainties}
\label{sec:sys_errors}

When comparing models to observations, we include
systematic uncertainties owing to cosmological parameter variations
and possible resolution limitations of the simulations.
In \cite{villasenor2021a}, we performed a study of the changes in the \Lya\ flux power spectrum $P(k)$ 
induced by small variations of the cosmological parameters within the constraints from \cite{Planck_collaboration_2020}. 
Our results suggested that 
uncertainties in the cosmological parameters could
cause a fractional change 
of $\lesssim 5\%$ on the hydrogen effective optical depth in the redshift range $2 \lesssim z \lesssim 5$ and a similar $\lesssim 5\%$ effect in $P(k)$ for scales 
$0.002 \lesssim k \lesssim 0.2 \,\, \mathrm{ s \, km^{-1}}$ and redshifts $2 \lesssim z \lesssim 5$. 
For this reason, we include here an additional systematic uncertainty 
$\sigma_{\mathrm{cosmo}}$ of $5\%$ to 
the observational determinations of the \Lya\ power spectrum.  
For the \HeII\ effective optical depth, we estimate similar variations of $\sim 5\%$ at  $2 \lesssim z \lesssim 3$ from 
differences in the mean baryonic density of different cosmologies.  We therefore include a $\sigma_{\mathrm{cosmo}} = 5\% $ to the measurements of \taueffHe as well.      
         
In Appendix \ref{sec:resolution} we present a resolution convergence study where we compare the forest flux power spectrum from three simulations performed with the same cosmological 
parameters and photoionization and photoheating histories.
The initial conditions used for the runs were generated to preserve the large-scale modes in common to each 
simulation, such that the properties of the simulations could be compared directly on shared spatial scales. The three simulations 
model
a box of size $L= 50 h^{-1}$Mpc 
and
$N=512^3$, $N=1024^3$, or
 $N=2048^3$ cells and particles.
The corresponding spatial resolutions are $\Delta x \simeq 98 \,h^{-1}$Mpc, 
$\Delta x \simeq 49 \,h^{-1}$Mpc, and $\Delta x \simeq 24 \,h^{-1}$Mpc, respectively.
In comparing the moderate resolution ($\Delta x \simeq 49 \,h^{-1}$Mpc) 
and high resolution ($\Delta x \simeq 24 \,h^{-1}$Mpc) simulations,
we measure small fractional differences $\Delta P(z,k) / P(z,k)$ of $\lesssim 5\%$ for the 
large scales $k \lesssim 0.02 \,\, \mathrm{s\, km^{-1}}$.
On small scales, $ 0.02 \lesssim k \lesssim 0.2 \,\, \mathrm{s\, km^{-1}}$, the fractional differences are 
slightly larger ($\lesssim 12\%$).  

To approximate resolution effects 
from the grid of simulations used for our analysis ($N=1024^3$, $\Delta x \simeq 49 \,h^{-1}$Mpc), we add an additional systematic uncertainty 
$\sigma_{\mathrm{res}}$ to the observational determinations of the flux power spectrum and the \HeII\ effective optical depth.
For $P(k)$, the additional uncertainty $\sigma_{\mathrm{res}}(z,k) = \Delta P(z,k)$
is set equal to the difference between $P(k)$ from the $N=1024^3$ and 
$N=2048^3$ reference simulations used for our resolution study.   
For the \HeII\ effective optical depth 
the impact of resolution is a small increase of $\lesssim 3\%$ from the $N=1024^3$ box to the
$N=2048^3$ run at $z<3$; we then add an uncertainty of $\sigma_{\mathrm{res}}(z) = 3\%$ to the estimate of \taueffHe. We note that the systematic errors added to \taueffHe\ are 
significantly smaller than the observational uncertainties    $\sigma_\mathrm{obs} \sim 12 - 45\%$ of \cite{Worseck+2019}.

The total uncertainty applied to the observational determinations of $P(k)$ and \taueffHe is finally given by the quadrature sum of the errors as

\begin{equation}
\sigma_\mathrm{total} \, = \, \sqrt{ \sigma_{\mathrm{obs}}^2 + \sigma_{\mathrm{cosmo}}^2 +  \sigma_{\mathrm{res}}^2 }
\label{eq:sigma_total}
\end{equation} 

\noindent
where $\sigma_{\mathrm{obs}}$ is the reported observational uncertainty in the flux power spectrum or helium opacity respectively.

In their study, \cite{wolfson2021} showed the importance of using the covariance matrix when inferring the
temperature of the IGM from measurements of the \Lya\ power spectrum and wavelet statistics.  
For our MCMC analysis we use the covariance matrices of $P(k)$ 
in the likelihood calculation (see \S\ref{sec:uvb_inference}). 
To reflect the increased uncertainty from Eq. \ref{eq:sigma_total}, we rescale the 
elements of the covariance matrices according to

\begin{equation}
\mathbf{C}[i,j] \, = \, \mathbf{C}_\mathrm{obs}[i,j] \frac{\sigma_{\mathrm{total},i} \,\, \sigma_{\mathrm{total},j}}{\sigma_{\mathrm{obs},i} \,\, \sigma_{\mathrm{obs},j}},
\label{eq:rescaled_covariance}
\end{equation} 
\noindent
where $\mathbf{C}_\mathrm{obs}$ is the reported covariance matrix of $P(k)$ taken from the published observational datasets
used for our analysis.

\begin{figure*}
\includegraphics[width=\textwidth]{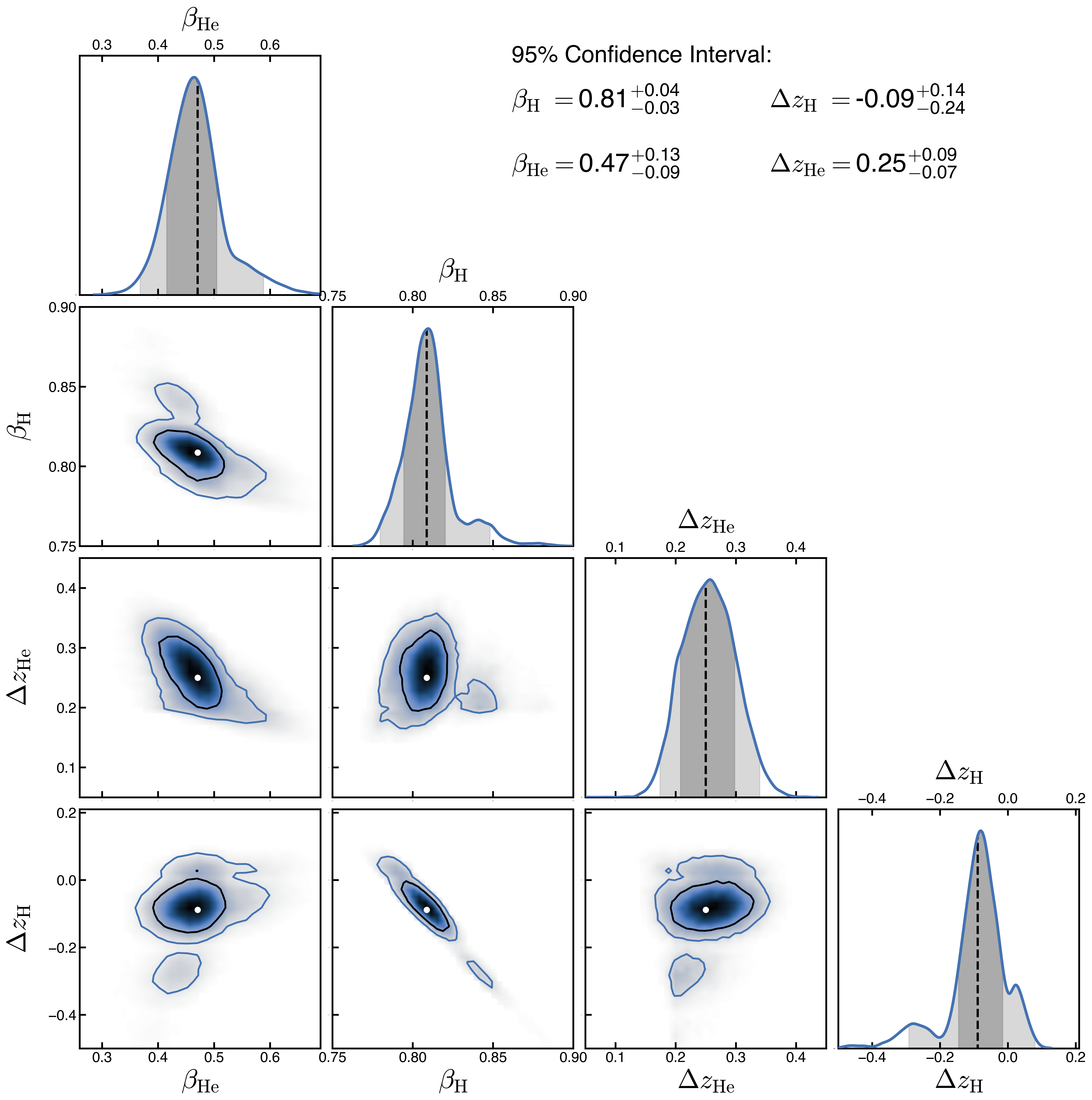}
\caption{Results from the Bayesian inference procedure, showing one- and two-dimensional projections of the posterior distributions
for the parameters $\theta= \{ \beta_{\mathrm{H}}, \,\Delta z_{\mathrm{H}}, \, \beta_{\mathrm{He}}, \,\Delta z_{\mathrm{He}}\}$.
The parameter constraints were obtained by 
fitting the observed flux power spectrum of the \Lya\ forest and the \HeII\ effective optical depth \citep{Chabanier+2019, boera2019a,  Worseck+2019, Irsic+2017a} with a grid of CHIPS simulations. 
The posterior distribution shows a clear global maximum, and while other local maxima are present their peak likelihoods 
are significantly lower than the global maximum.    
The resulting best-fit parameters and their 95\% confidence intervals are shown in the top right corner.}    
\label{fig:corner}
\end{figure*}

\subsection{Inference of the UVB Model}\label{sec:uvb_inference}

To find the photoionization and photoheating rates that best reproduce the properties of the IGM encoded in the 
flux power spectrum of the \Lya\ forest $P(k)$ and the helium effective optical depth \taueffHe, we apply an MCMC
sampler to compare the simulated $P(k)$ and \taueffHe to the observational measurements over the redshift  
and frequency range where data are available. The likelihood function for the model given by the parameters  
$\theta= \{ \beta_{\mathrm{H}}, \,\Delta z_{\mathrm{H}}, \, \beta_{\mathrm{He}}, \,\Delta z_{\mathrm{He}} \} $  is evaluated as:

\begin{equation}
\begin{aligned}
& \ln \mathcal{L}(\theta) =  
- \frac{1}{2} \sum_z \left[ \left( \frac{  \tau_{\mathrm{obs}}(z) - \tau(z|\theta) }{ \sigma_\tau(z)  } \right) ^2 + \ln \big( 2\pi \sigma_\tau(z)^2 \big)  \right] \\
& - \frac{1}{2}  \sum_\mathrm{datasets} \sum_z \left[  \mathbf{\Delta}^{T} \mathbf{C}^{-1} \mathbf{\Delta} +  \ln \operatorname{det}(\mathbf{C}) + N \ln 2 \pi  \right],
\end{aligned}
\label{eq:mcmc_likelihood}
\end{equation}
\noindent
where the first term compares the \HeII\ effective optical depth measured from our simulations  $\tau(z|\theta)$ 
for a given photoionization and photoheating model represented by the vector $\theta$ to the observational measurement $ \tau_{\mathrm{obs}}(z)$ 
from \cite{Worseck+2019} with total (observational + systematic) uncertainty $\sigma_\tau(z)$.  The second term compares the \Lya\ power 
spectrum, with $\mathbf{\Delta}$ denoting the difference vector between the observations and the model 
$\mathbf{\Delta} = P_{\mathrm{obs}}(z,k) - P(z,k|\theta)$. Here, $\mathbf{C }$ corresponds to the covariance matrix of size 
$N \times N$ associated with the observational determination, where $N$ is the number of points of each measurement. 
To compute $P(z,k|\theta)$ and $\tau(z|\theta)$ for arbitrary values of the parameters $\theta$ not directly
simulated by our grid, we perform a four-dimensional linear interpolation of the sixteen neighboring simulations in parameter space.

As described in \S\ref{sec:observational_data}, we employ the datasets from \cite{Chabanier+2019} ($2.2 \leq z \leq 4.6$),  
\cite{Irsic+2017a} ($3.0 \leq z \leq 4.2$),
and \cite{boera2019a} ($4.2 \leq z \leq 5.0$)  for the observational measurements of the power spectrum used in our analysis.
While there is some overlap in the measurements from the datasets,  in general their determinations are consistent with each other.
For this reason we include all the data points 
from each dataset for the likelihood calculation.  The only significant discrepancy is at $z=4.6$ where $P(k)$ from \cite{Chabanier+2019} is lower 
to the determination from \cite{boera2019a}. We repeated our the analysis excluding the $z=4.6$ measurement from \cite{Chabanier+2019} and 
obtained similar posterior distributions. We conclude that this difference does not impact our result. 

The contribution from each redshift bin to the total log likelihood $\ln \mathcal{L}$ (Eq. \ref{eq:mcmc_likelihood}) from $P(k)$ and
\taueffHe in our analysis is presented in Table \ref{tab:likelihood}.
The quantity $\Delta \ln{\mathcal{L}}$ is evaluated as the first and second terms
of Eq. \ref{eq:mcmc_likelihood} for \taueffHe and $P(k)$, respectively, for each redshift bin.
The power spectrum mostly strongly influences the log likelihood, with data from redshifts $z=2.4$ and $z=4.2$ 
inducing the largest fractional changes in the likelihood.

\begin{deluxetable}{ccc|ccc}[h]
\label{tab:likelihood}
\tablenum{2}
\caption{Redshift bin contribution to the Likelihood \label{tab:obs_datasets}}
\tablewidth{\columnwidth}
\tablehead{
	\colhead{Type} & \colhead{$z$} & \colhead{$-\Delta \ln{\mathcal{L}}$} & \colhead{Type} & \colhead{$z$} & \colhead{$-\Delta \ln{\mathcal{L}}$}\\[-16pt]
}
\startdata
$P(k)$ & 2.2 & 330.6 & $P(k)$ & 4.2 & 489.2   \\
$P(k)$ & 2.4 & 363.3 & $P(k)$ & 4.4 & 135.7   \\
$P(k)$ & 2.6 & 229.2 & $P(k)$ & 4.6 & 190.2   \\
$P(k)$ & 2.8 & 297.0 & $P(k)$ & 5.0 & 40.0   \\
$P(k)$ & 3.0 & 215.1 & \taueffHe  &  2.30   & 0.5   \\
$P(k)$ & 3.2 & 134.4 & \taueffHe  &  2.54   & 0.2   \\
$P(k)$ & 3.4 & 113.8 & \taueffHe  &  2.66   & 0.3   \\
$P(k)$ & 3.6 & 84.1  & \taueffHe  & 2.74 & 1.0   \\
$P(k)$ & 3.8 & 137.1 & \taueffHe  & 2.82 & 2.3   \\
$P(k)$ & 4.0 & 180.3 &            &     &   \\
\enddata
\vspace{-6mm}
\end{deluxetable}

The covariance matrices of $P(k)$ are taken from the published observations.  
We note that \cite{Irsic+2017a} provides the complete covariance of $P(k)$ across the seven redshift bins of their measurement. For this 
dataset we employ the reported full covariance and the residual vector $\mathbf{\Delta}$ consists of the $P(k)$ difference from the model and
observation concatenated over the seven redshift bins.    

While our likelihood 
analysis uses the reported covariance matrices from the observations, in Appendix \ref{sec:covariance_matrices} we present the 
covariance of $P(k)$ measured from a subset of our simulations to quantify the differences induced by variation of our four model 
parameters. We show that the structure of the covariance is mantained across our simulations and we measure 
relatively small variations between 
the different models.       

We emphasize that our approach differs from previous studies of 
the thermal history of the IGM 
\citep[e.g.,][]{ bolton2014a, Nasir+2016, hiss2018a,  boera2019a,   walther2019a, gaikwad2020b} in an important aspect. Typically,
the method  adopted to infer the thermal state of the IGM from observations of the \Lya\ forest involves marginalizing over the thermal
parameters $T_0$ and $\gamma$ in the approximate power-law density-temperature relation \citep{hui1997a} $T(\Delta) = T_0 \Delta^{\gamma-1}$, where 
$\Delta = \rho_\mathrm{gas}/\bar{\rho}$ is the gas overdensity.  This marginalization is often performed independently for each redshift.
Instead, our approach to find the optimal photoionization and photoheating rates that best reproduce the observational measurements is to compare the
simulated $P(k)$ and \taueffHe to the observations over the full redshift range where data is available, namely $2.2 \leq z \leq 5.0$ for 
$P(k)$ and $2.2 < z < 3.0$ for \taueffHen. 

In our approach, the performance for a given UVB model to match the observations is  evaluated over the complete
self-consistently evolved reionization and thermal history of the IGM that results from that model. Since the properties 
of the gas at one redshift cannot be disentangled from its properties at previous epochs, the thermal and ionization 
structure of the forest depends on the time-dependent photoheating and photoionization rate.
Both $T_0$ and $\gamma$ evolve along continuous trajectories with redshift, and we therefore 
marginalize  over the full simulated histories of IGM properties.

Our simulations span a wide range of reionization histories for hydrogen in the IGM. 
Instead of following the common practice of rescaling the optical  depth of the simulated skewers in post-processing to 
match the observed mean transmission of the forest, our method self-consistently follows the ionization evolution of 
hydrogen and the effective optical depth \taueffH encoded in the redshift-dependent power spectrum of the transmitted flux.
Furthermore, during our inference procedure, we do not assume a power-law 
approximation for the density-temperature distribution of IGM gas or
apply a post-processing procedure  that artificially modifies the temperature of the gas in the simulations. 
Instead, our synthetic \Lya\ spectra reflect the real $\rho_\mathrm{gas}-T$ distribution from the simulations.
This improvement proves relevant, as we find that a single power law is not a good fit over the full range of 
gas densities responsible for the bulk of the  \Lya\ absorption signal (see Appendix \ref{sec:phase_diagram}).

The posterior distribution for our parameters $\theta= \{ \beta_{\mathrm{H}}, \,\Delta z_{\mathrm{H}}, \, \beta_{\mathrm{He}}, \,\Delta z_{\mathrm{He}} \} $
resulting from the Bayesian inference procedure is shown in Figure \ref{fig:corner}. 
A clear global maximum of the 
posterior distribution is observed, and while the posterior shows other local maxima their likelihoods are significantly 
lower than the global peak.   
The four model parameters are well constrained and show only small correlations that arise  
from the weak degeneracies in the resulting ionization and thermal histories produced by the different photoionization and photoheating rates. 
Our best-fit parameters and their 95\% confidence limits are

\begin{equation}
\begin{aligned}
\beta_{\mathrm{H}} \,  &= \, 0.81^{+0.04}_{-0.03}  \,\,\,\,\,\,\,\,\,\,  \Delta z_{\mathrm{H}}  \,\, \,=\, -0.09^{+0.14}_{-0.24}\\ 
\beta_{\mathrm{He}} \, &= \, 0.47^{+0.13}_{-0.09}  \,\,\,\,\,\,\,\,\,\, \Delta z_{\mathrm{He}} \, =\,  0.25^{+0.09}_{-0.07}.
\end{aligned}
\label{eq:parameter_constraints}
\end{equation}    
%
To measure the properties of the IGM that result from our best-fit distribution, we sample $P(k)$, \taueffHn, and \taueffHen, together with the thermal parameters $T_0$ and $\gamma$, over the posterior distribution of the parameter vector $\theta$, 
resulting in determinations of the highest-likelihood and 95\% confidence interval for the forest statistics and thermal history. When necessary, we interpolate results for values of 
$\theta$ not directly simulated by our grid.

\begin{figure*}
\includegraphics[width=\textwidth]{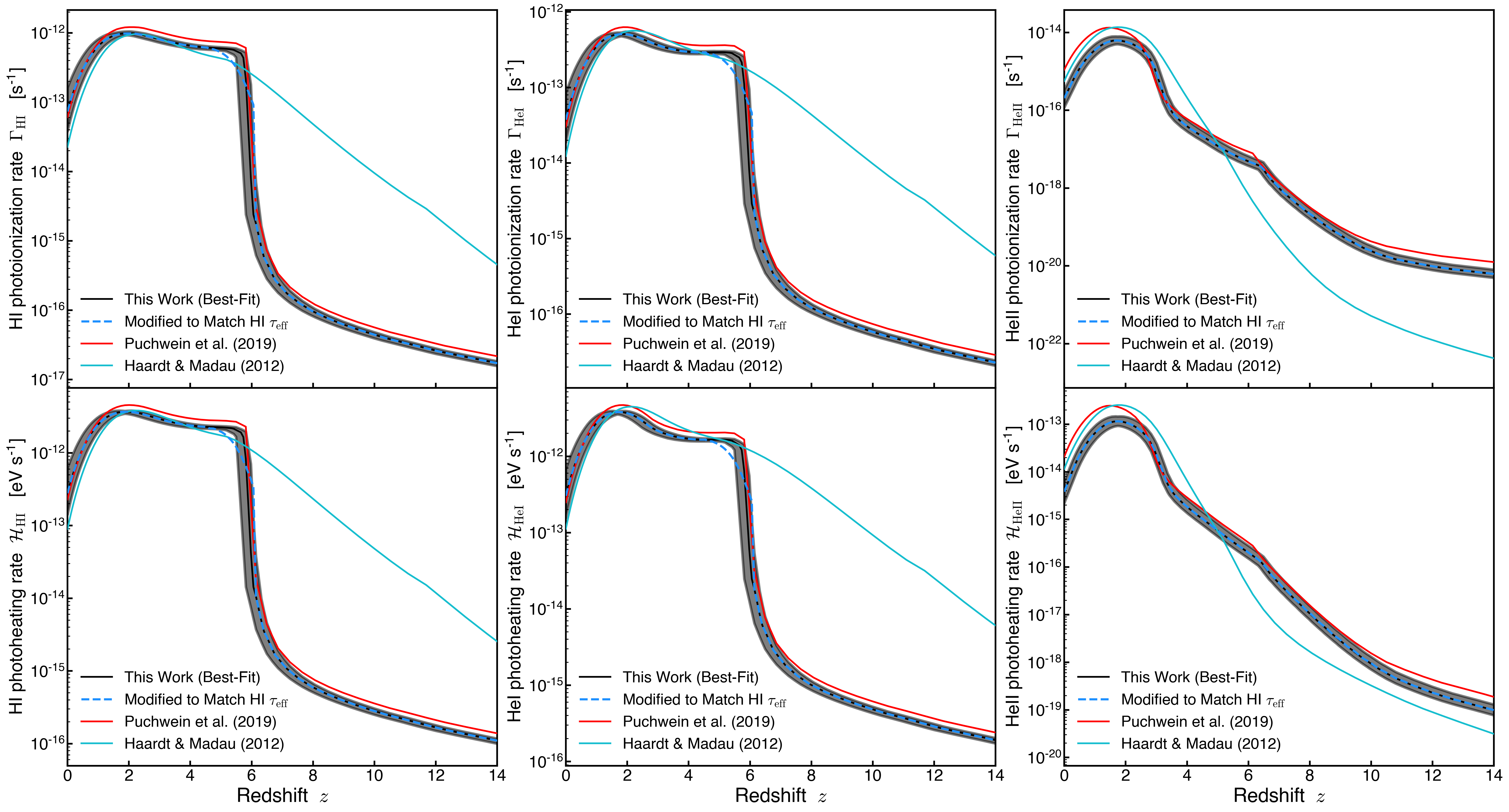}
\caption{Best-Fit (black lines) and 95\% confidence intervals (grey bands) for the photoionization ($\Gamma$, top) and photoheating ($\cal{H}$, bottom) rates for neutral 
hydrogen (\HI, left), neutral helium (\HeI, center), and singly ionized helium (\HeII, right) obtained from our MCMC analysis.
The \textit{modified} \HI\ and \HeI\ photoionization 
and photoheating rates (dashed blue lines) are identical to the reference best-fit model except for the redshift range $ 4.8 \leq z \leq 6.1$ where they have been modified to 
produce an evolution of the hydrogen effective optical depth consistent with the observational determinations of \cite{Bosman_2018} for $z>5$ 
(see \S \ref{sec:evolution_tau_HI } and \S \ref{sec:Gamma_HI } for details).
For reference, we also show the models from \cite{puchwein2019a} (red) and \cite{haardt2012a} (cyan). }
\label{fig:uvb_result}
\end{figure*}

\begin{figure*}
\includegraphics[width=\textwidth]{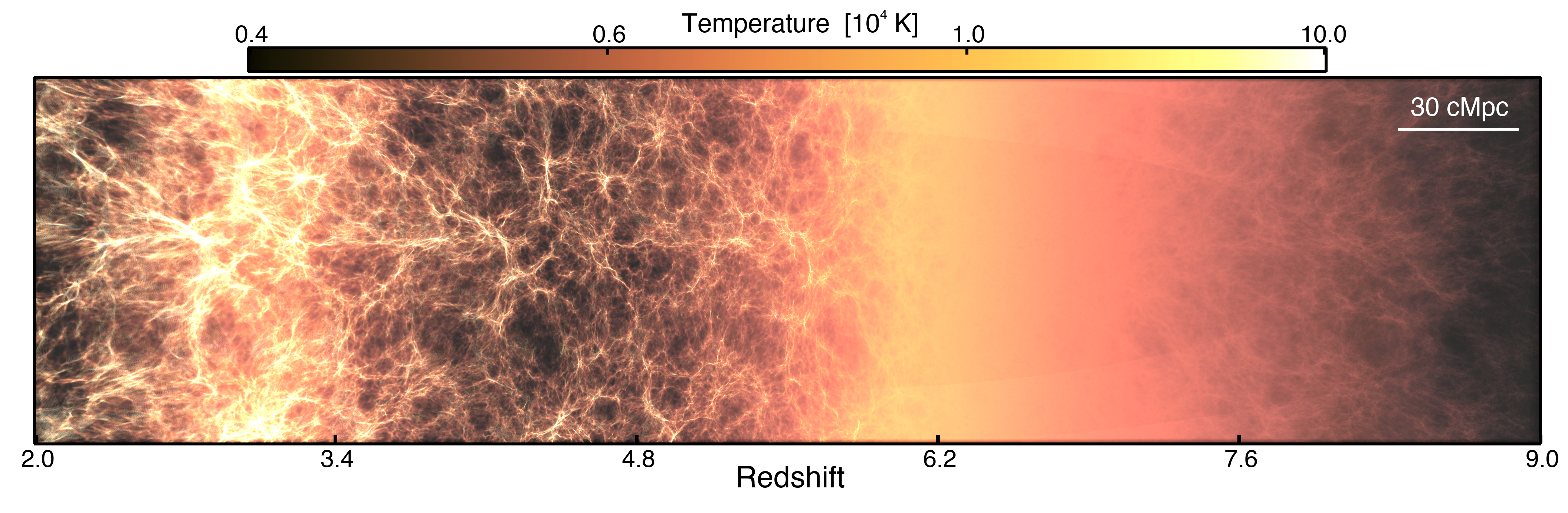}
\caption{ Redshift evolution of the gas temperature from a high-resolution simulation ($L=50\,h^{-1}$Mpc, $N=2048^3$ cells and particles) that employed our best-fit model 
for the photoheating and photoionization rates. The image displays the monotonic increase in the temperature of the IGM due to hydrogen reionization for $z \gtrsim 6.0$ followed 
by an epoch of cooling of the IGM due to cosmic expansion.
The onset of helium reionization ($z \sim 4.5$) initiates a second epoch of heating of the IGM that ends at $z\sim 3$ when \HeII\ reionization completes.
A second epoch of cooling due to cosmic expansion then follows.
The temperature increase of gas collapsing into the filamentary cosmic web as large-scale structure develops is also visible in the image.}
\label{fig:temperature}
\end{figure*}

\section{Results and Discussion}
\label{sec:discussion}

By comparing the flux power spectrum and the
\HeII\ effective opacity in our \Sim simulation grid to
observational determinations, we can infer a set of photoionization and photoheating histories that, when input in cosmological hydrodynamical simulations, result in statistical properties of the \Lya\ forest that are consistent with observations. 
In this section, we present the best-fit rates obtained from our inference procedure, as well as the \Lya\ forest statistics 
and thermal evolution of the IGM produced by our best-fit UVB model.
We compare our results to previous work and finalize our discussion by describing the limitations of our method. 

\subsection{Best-Fit Photoionization and Photoheating Rates}
\label{sec:result_uvb_rates}

Figure \ref{fig:uvb_result} shows our  best-fit model for the photoionization and photoheating rates along with the corresponding 95\% confidence 
interval that results from our MCMC marginalization of the UVB rates over the posterior distribution of the model parameters obtained from  our MCMC analysis. We note that the 
transformations applied in this work  to generate new photoionization and photoheating rates from the reference model \citep{puchwein2019a} are relatively simple and preserve the functional form of 
the P19 model. While we allow for orders of magnitude variations in the rates,
the flexibility of the ionization and thermal histories sampled here is limited by the fixed shape of the UVB model employed in our simulation grid. A study that allows for more flexibility 
in the photoionization and photoheating rates of hydrogen and helium will be the scope of future work.  

\subsection{$P(k)$ Model Comparison with the Data} \label{sec:results_model}

Figure \ref{fig:power_spectrum_all} shows the evolution of the
best-fit flux power spectrum and 95\% confidence intervals
over the redshift range $2.2 \leq z \leq 5.0$ that result from marginalizing $P(k)$ over the posterior
distribution of model parameters
$\theta= \{ \beta_{\mathrm{H}},\,\Delta z_{\mathrm{H}},\, \beta_{\mathrm{He}},\,\Delta z_{\mathrm{He}}\}$.
Our best-fit synthetic power spectrum shows
good agreement with the large scale $P(k)$ measured by the eBOSS experiment \citep{Chabanier+2019} 
in the range $2.4 \lesssim z \lesssim 4.2$, suggesting that the mean transmission \Fmean of the forest
inferred by our analysis is consistent with the measurements by \citet{Chabanier+2019}.
Only for $z=2.2$ and $z=4.4 - 4.6$ do our
results show significant differences with the eBOSS data set. 
At $z=2.2$, the $P(k)$ from eBOSS is higher than our results  by $\sim 8 - 20 \%$ 
on scales $0.008 \lesssim k \lesssim 0.02 \,\, \mathrm{s\,km^{-1}}$. 
This modest tension may suggest that the hydrogen opacity \taueffH
is underestimated by $\sim 10 \%$ in our modeling
relative to eBOSS.
At $z=4.4$ and $z=4.6$ the opposite is true, and
our best-fit $P(k)$ on large scales is $\sim 15\%$ and $\sim 25\%$ higher than the eBOSS measurements, respectively.
These small discrepancies could be alleviated, e.g., by a small 15\% decrease of the \HI\ photoionization rate at $z=2.2$ and by a comparable small increase in the same quantity at $z=4.4 - 4.6$ by $\sim 10 - 20\%$.

Our results also agree on large and intermediate scales 
($ 0.003 \lesssim k \lesssim 0.06 \,\, \mathrm{s \, km^{-1}}$) with the estimates of \cite{Irsic+2017a}. 
The best-fit model reproduces the turnover in the observed dimensionless power spectrum 
$\Delta^2(k)=\pi^{-1}kP(k)$ at  $k\sim 0.02-0.03 \,\, \mathrm{s \, km^{-1}}$, 
and generally lies within the observational uncertainties at intermediate scales 
$ 0.01 \lesssim k \lesssim 0.06  \,\, \mathrm{s \, km^{-1}}$. 
Only at redshift $z=3.4$ and $z=3.8$ the $P(k)$ measurements show some differences relative to the model. 
At $z=3.4$ the data are higher than the model by $\sim 5-20\%$.
A similar discrepancy is observed when comparing 
\cite{Irsic+2017a} with the determinations by eBOSS at the same redshift, suggestive
of a slightly higher \HI\ opacity \taueffH in the former sample.
Differences with the model are more significant at $z=3.8$, where
on intermediate scales ($k \gtrsim 0.2 \,\, \mathrm{s \, km^{-1}}$) the measurements of 
\cite{Irsic+2017a} are lower than the model by $\sim 10-20\%$,  while on large scales ($k \lesssim 0.2 \,\, \mathrm{s \, km^{-1}}$) their estimates are higher than both the 
model and the determinations by eBOSS by $\sim 5 - 30\%$.

Our model is in good agreement with the high-redshift measurements of $P(k)$ 
by \cite{boera2019a}, with minor differences that could be addressed by small modifications to the early photoheating history. At $z=4.2$, $z=4.6$, and  $z=5.0$, 
our best-fit $P(k)$ is consistent with their data points on large scales $k\lesssim 0.02 \,\, \mathrm{s \, km^{-1}}$, suggesting that our inferred IGM \HI\ opacity matches that measured by \cite{boera2019a}.
The model also reproduces the cutoff in $\Delta^2(k)$ at $k\sim 0.02-0.03\,\, \mathrm{s \, km^{-1}}$ and the consistency with the observations extends to small scales $ k \lesssim 0.1 \,\, \mathrm{s \, km^{-1}}$.
Discrepancies appear only on the smallest scales $ 0.1 \lesssim k \lesssim 0.2 \,\, \mathrm{s \, km^{-1}}$ where the model has 
less power ($\sim 10-30\%$) than \cite{boera2019a}.
This may suggest that the temperature of the IGM has been overestimated by the 
model in the redshift range $4 \lesssim z \lesssim 5$ (see \S \ref{sec:evolution_temperature } for a discussion of this issue).

\begin{figure*}
\includegraphics[width=\textwidth]{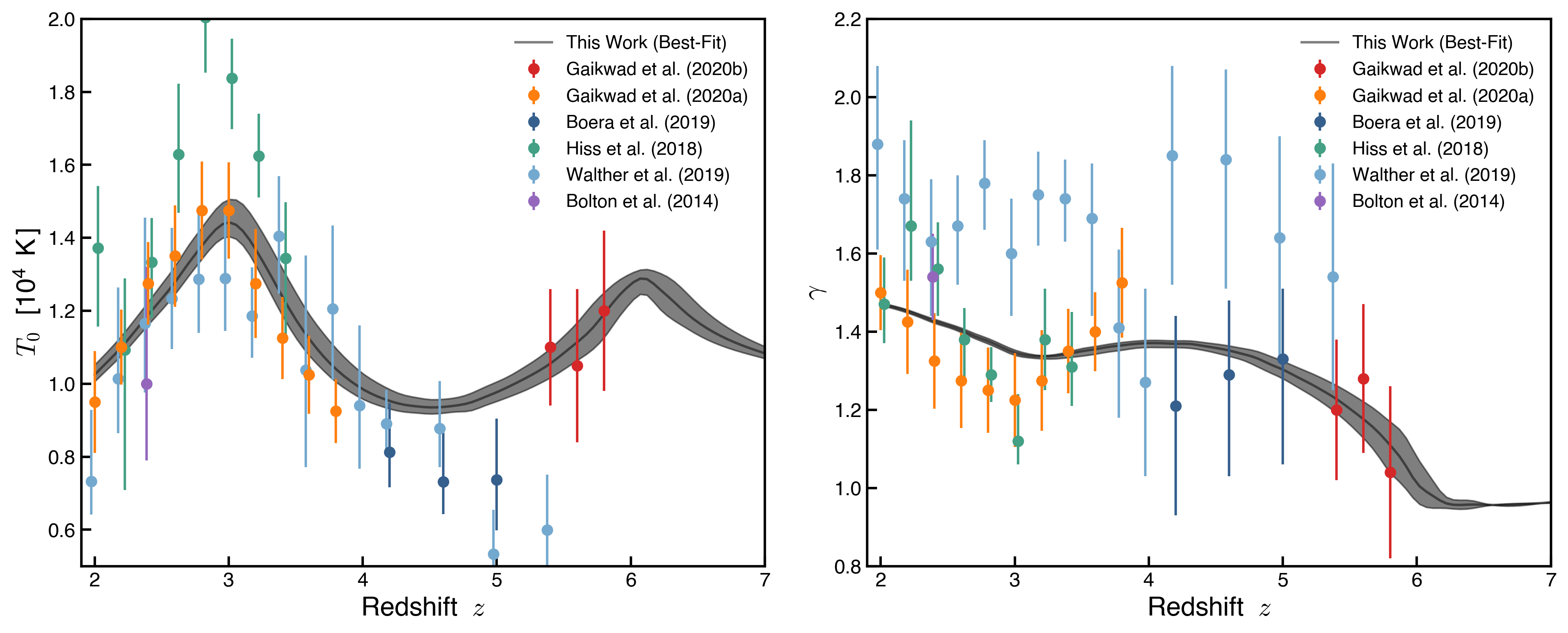}
\caption{Redshift evolution of the parameters $T_0$ and $\gamma$ [Eq. (\ref{eq:rho_T_relation})] from the best-fit model (black lines) and 95\% confidence interval (gray band) obtained from our MCMC analysis.
The data points show the values of  $T_0$ and $\gamma$ inferred 
from observations of the \Lya\ forest by 
\citet{bolton2014a,hiss2018a,boera2019a,walther2019a, gaikwad2020a,gaikwad2020b}.
Our results reveal two peaks in the evolution of $T_0$ due to hydrogen reionization at $z\sim 6$ and helium reionization at $z\sim 3$, and are consistent with previous measurements from 
\cite{gaikwad2020a, gaikwad2020b}.}
\label{fig:T0_gamma}
\end{figure*}

\subsection{Evolution of the IGM Temperature}
\label{sec:evolution_temperature }

The flux power spectrum and helium opacity tightly constrain the time-dependent photoionization and photoheating rates,
which in turn determine the IGM ionization and thermal history.
The redshift evolution of the gas temperature is illustrated in Figure \ref{fig:temperature}
which is generated from a slice through a high-resolution simulation ($L=50\,h^{-1}$Mpc, $N=2048^3$ cells and particles)
using  our best-fit photoionization and photoheating rates.
The figure shows the monotonic increase in the temperature of the IGM during hydrogen reionization at $z \gtrsim 6$. After hydrogen reionization
completes by $z\sim 6$, the input of energy into the IGM falls dramatically, and the gas then cools primarily through adiabatic expansion.
This first epoch of IGM cooling lasts 
until the onset of helium reionization ($z\sim 4.5$) when extreme UV radiation from AGNs ionizes \HeII\ atoms and drives a second epoch of reheating that completes by $z\sim 3$ and is followed by a second epoch of adiabatic cooling.

The thermal state of diffuse IGM gas is often modeled with the power-law relation 
\citep{hui1997a, puchwein2015a, mcquinn2016a}

\begin{equation}
T(\Delta) = T_0 \Delta ^{\gamma - 1}.
\label{eq:rho_T_relation}
\end{equation} 
\noindent
We fit the power law relation to the gas density-temperature 
distribution in each of the simulations from the CHIPS grid and at multiple epochs, $2\leq z \leq 9$, following the procedure presented in \cite{villasenor2021a}. We restrict the fit to the overdensity range $0 \leq \log_{10} \Delta \leq 1$, as we find 
that in our simulations a single power law does not accurately describe the wider range 
$-1 \leq \log_{10} \Delta \leq 1$ (see Appendix \ref{sec:phase_diagram}).

Figure \ref{fig:T0_gamma} shows the redshift evolution of the parameters $T_0$ and $\gamma$ from 
our best-fit model and the 95\% confidence interval
that results from our  MCMC marginalization over the posterior distribution of the photoionization and photoheating rates. For comparison, we also depict the data points for these parameters
inferred from the properties of the \Lya\ forest by
\cite{bolton2014a}, \cite{hiss2018a}, \cite{boera2019a},  \cite{walther2019a},
\cite{gaikwad2020a}, and \cite{gaikwad2020b}.
        
The inference from \cite{boera2019a} and \cite{walther2019a} follow similar methodologies.
They generate flux power spectra from simulations run with different thermal 
histories, resulting in multiples trajectories for the evolution of $T_0$ and $\gamma$.
For each redshift bin they determine the best-fit $T_0$, $\gamma$, and mean transmitted 
flux \Fmean by performing Bayesian inference and comparing the simulated flux power spectra to observations of the \Lya\ forest $P(k)$.    
\cite{bolton2014a}  and \cite{hiss2018a} measure a set of values for the Doppler parameter $b$ and \HI\ column density $N_{\mathrm{HI}}$ directly from the forest by
decomposing the absorption spectra into a collection of Voigt profiles. They infer the parameters $T_0$ and $\gamma$ by comparing simulations with different $b-N_{\mathrm{HI}}$ distributions 
to the observed one. \cite{gaikwad2020a} follow a similar approach by comparing simulated \Lya\ forest spectra to Voigt profiles fitted to the observed transmission spikes in the
inverse transmitted flux $1-F$ at $z>5$.
\cite{gaikwad2020b} report more precise determinations by inferring $T_0$ and $\gamma$ from the combined constraints obtained
through a comparison of simulated \Lya\ forest absorption with the observed flux power spectra, $b-N_{\mathrm{HI}}$ distributions, wavelet statistics, and curvature statistics.

As shown in Figure \ref{fig:T0_gamma}, the temperature evolution from our best-fit model presents a
first peak ($T_0 \simeq 1.3 \times 10^4 \, \mathrm{K}$)  at the end of hydrogen 
reionization ($z\sim 6.0$) followed by an epoch of adiabatic cooling from cosmic expansion.
Our results agree well with the high redshift measurements of $T_0$ and $\gamma$ at $ 5.4 \leq z \leq 5.8$ from \cite{gaikwad2020a}. We note that their estimates also suggest a period of cooling at these epochs, and from their result it is possible to infer a peak in $T_0$ from H reionization sometime at redshift $z\gtrsim 5.8$.

In our model,
the IGM continues to cool until the onset of helium reionization, 
and the temperature reaches a local minimum of 
$T_0(z\sim 4.5) \simeq 9.5\times 10^3\, \mathrm{K}$.
Evidence of this transition can also be seen in the measurements from \cite{boera2019a},
where $T_0$ shows little evolution from 
$z=5.0$ to $z=4.6$ and then a slight increase to $z=4.2$.
Nevertheless,  there are significant differences between $T_0$ from the model at $4 \lesssim z \lesssim 5$  and the 
measurements from \cite{boera2019a},
as the temperature predicted by our model is higher than their inferred values of 
$T_0 \sim 7.4 \times 10^3 \,\mathrm{K}$ and $T_0 \sim 8.1 \times 10^3 \,\mathrm{K}$ at $z=4.6-5$ and $z=4.2$, respectively.
The higher temperatures in our model 
reflect a suppressed power spectrum of the \Lya\ flux on small scales ($0.1 \lesssim k \lesssim 0.2 \,\,\mathrm{s \, km^{-1}}$) compared to the $P(k)$ measurement from 
\cite{boera2019a} at $4.2 \leq z \leq 5.0$ (see Figure \ref{fig:power_spectrum_all}).
Decreasing the photoheating from the UVB during $z\gtrsim 4$ would decrease 
the temperature of the IGM at this epoch and potentially alleviate this discrepancy. 

In Appendix \ref{sec:colder_IGM} we present scenarios were the mid-redshift IGM is set to be colder
compared to our model by decreasing the best-fit \HI\ and \HeI\ photoheating rates at $4.2 \leq z \leq 6.2$. We find that reducing $\mathcal{H}_\mathrm{HI}$ and 
$\mathcal{H}_\mathrm{HeI}$ by $\sim 80\%$ at $z\sim6$ decreases the IGM temperature $T_0$ by $\sim  20\%$ making it consistent with the estimates from 
\cite{boera2019a} at $4.2 \leq z \leq 5.0$ with minimal impact in $T_0$ at $z \lesssim 3.5$ (see Figure \ref{fig:reduced_heating}).
Nevertheless,  we find that such colder evolution of $T_0$ is in conflict with the $z\sim 5.4$ 
estimate from \cite{gaikwad2020a} (see Figure \ref{fig:reduced_heating}).
This conflict indicates some degree of tension between the higher 
$T_0=1.10 \pm 0.16 \times 10^4 \, \mathrm{K}$ at $z\sim5.4$ from \cite{gaikwad2020a} and the 
low $T_0=7.37^{+1.13}_{-1.39} \times 10^3 \, \mathrm{K}$ at $z\sim 5.0$ from \cite{boera2019a}.      
 
After $z\sim 4.5$, radiation from AGN ionizes \HeII\ atoms in the Universe
and heats the IGM for a second time.
Our model predicts that $T_0$ increases monotonically until \HeII\
reionization completes at $z\sim 3$, resulting in a second peak in the temperature ($T_0 \simeq 1.4\times 10^4\, \mathrm{K}$) followed by a second epoch of cooling due to cosmic
expansion.
Our results for the evolution of $T_0$ during $z\lesssim 4.5$ are
consistent with the determinations from \cite{gaikwad2020b} and \cite{walther2019a} that
show a similar
$T_0$ history within the uncertainties during and after \HeII\ reionization,
as both show a peak in $T_0$ at $z\sim 2.8 - 3.0$.
Our $T_0(z)$ results are higher yet
consistent within the uncertainties from the measurement by \cite{bolton2014a} at $z=2.4$. 
The results presented by \cite{hiss2018a} also show the effects of 
\HeII\ reionization on the temperature of the IGM in the form a peak in the temperature at $z\sim 2.8$,
but their  peak value of $T_0\sim 2\times 10^4\, \mathrm{K}$ is significantly 
higher than our result and the measurements from \cite{gaikwad2020a} and \cite{walther2019a}.

The right panel of Figure \ref{fig:T0_gamma} shows our result for the evolution of the
density-temperature power-law index $\gamma$
(black line and shaded 95\% confidence interval).
At the end of hydrogen reionization, 
the gas in the IGM is mostly isothermal ($\gamma \sim 1$).
As the IGM cools and the 
low-density gas cools more efficiently, the index $\gamma$ increases in the interval 
$ 4.5 \lesssim z \lesssim 6$. 
During the reheating of the IGM from \HeII\ reionization, low-density gas heats faster
and $\gamma$ decreases until helium reionization completes. 
After helium reionization cooling from cosmic expansion
causes an increase on $\gamma$ for a second time.

The evolution of the power-law index in our model is consistent with
measurements from \cite{hiss2018a}, \cite{boera2019a},  \cite{gaikwad2020a},  and \cite{gaikwad2020b},
and shows deviations only for a few redshift bins after \HeII\ reionization
completes.
The transition in $\gamma$ after \HeII\ reionization in our model is not as pronounced as the determinations
from \cite{gaikwad2020b} and \cite{hiss2018a}. 

The results from \cite{walther2019a} 
show significantly higher values of $\gamma$ compared to all the other measurements. We have evaluated the plausibility
of a steep density-temperature relation ($\gamma > 1.6$) by simulating the extreme case in which all photoheating and photoionization
from the UVB stops after hydrogen reionization completes,  i.e. $\Gamma = 0$ and $\mathcal{H}=0$ for $z > 6$. We find that in the absence of 
external heating, as the IGM cools by adiabatic expansion, the overdensities cool down at a slower rate from compression by gravitational collapse. Here 
$\gamma$ tends to increase with decreasing redshift at a 
roughly constant rate of $\Delta \gamma / |\Delta z| \sim 0.18 $. Starting from an 
isothermal distribution of the gas in the IGM when H reionization finishes ($\gamma = 1$),  it takes a change in redshift  $ |\Delta z| \sim 3 - 3.5 $ 
for the gas distribution to steepen to $\gamma \sim 1.6$. 
Hence, we can reproduce values of 
$\gamma > 1.6$ at $z\sim 5$ only if hydrogen reionization completes very early at $z > 8$.

\begin{figure}
\includegraphics[width=0.47\textwidth]{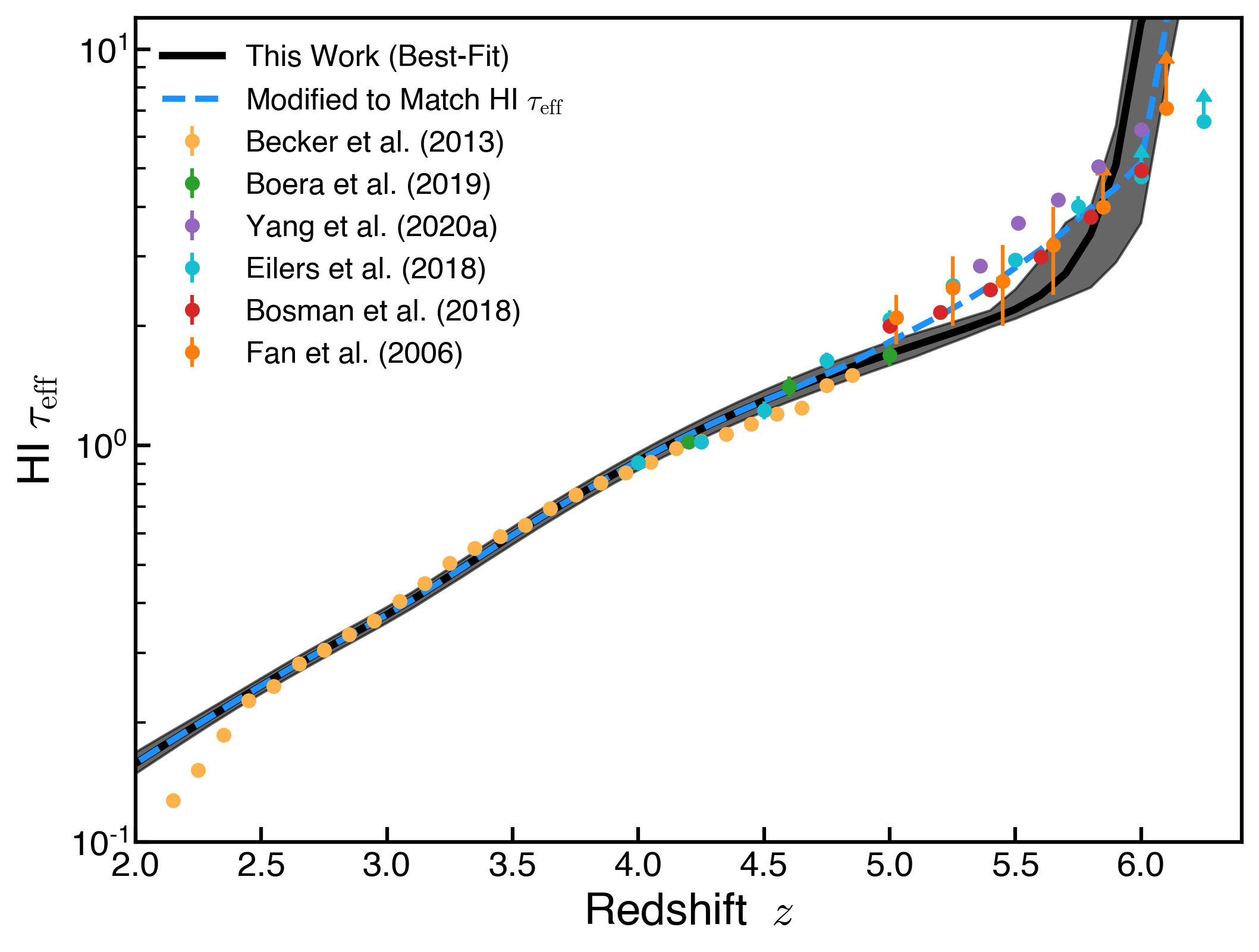}
\caption{  Redshift evolution of the hydrogen effective optical depth \taueffH from our best-fit
determination of the photoheating and photoionization rates (black line) and the
corresponding 95\% confidence interval. Data points show the observational measurements of \taueff from \cite{Fan+2006}, \cite{Becker+2013a}, 
\cite{Bosman_2018}, \cite{eilers2018a}, \cite{boera2019a}, and \cite{Yang+2020b}.
The model results show consistency
with the measurement from \cite{Becker+2013a} (yellow) for $2.5 \lesssim z \lesssim 4.2$ and are in 
good agreement with the determination from \cite{boera2019a} (green) for $4.2 \lesssim z \lesssim 5.0$.
At high redshift ($z>5$) the results from \cite{Yang+2020b} lie significantly 
higher than those from \cite{eilers2018a} and \cite{Bosman_2018} by $\sim 10-30 \%$.
In the redshift range $5 \lesssim z \lesssim 5.8$, the model shows lower \taueffH 
compared with the observations.
By modifying the best-fit  \HI\ photoionization rate $\Gamma_\mathrm{HI}$
as shown in \S \ref{sec:modified_uvb_model}, we can obtain a high-$z$
evolution of \taueffH (dashed blue) consistent
with the measurement from \cite{Bosman_2018} and \cite{Fan+2006}.    }     
\label{fig:tau_HI}
\end{figure}

\subsection{Evolution of the Hydrogen Effective Optical Depth} \label{sec:evolution_tau_HI }

The \HI\ effective optical depth \taueffH$=-\ln \langle F \rangle$ measured from the \Lya\ forest reflects the overall \HI\ content of the gas in the IGM. Hence, \taueffH probes the ionization 
state of hydrogen in the medium and can be used to constrain the intensity of the ionizing UVB.   
In our work, constraints obtained for the \HI\ photoionization rate $\Gamma_{\mathrm{HI}}$
derive from the power spectrum of the \Lya\ transmitted flux itself as we do not include 
the observational determinations of \taueffH as constraints in our inference procedure.

The power spectrum $P(k)$ of the flux fluctuations [Eq. (\ref{eq:delta_F})] is itself sensitive to the 
hydrogen effective optical depth.  Because of the non-linear relation $F = \exp(-\tau)$, 
the normalization of $P(k)$ on most scales relevant to this work ($0.002 \lesssim k \lesssim 0.1 \,\, \mathrm{s\, km^{-1}}$) is 
affected by the value of \taueffH obtained from the skewer sample used for the measurement. 
Thus, including the
effective optical depth of the forest does not provide additional independent information for constraining the model.
See Appendix \ref{sec:rescale_tau} for a discussion on the impact 
that \HI\ \taueff has on the \Lya\ flux power spectrum.

Figure \ref{fig:tau_HI} shows the redshift dependence of \taueffH from our best-fit determination
of the photoheating and photoionization rates (black line) and the
corresponding  95\% confidence interval.
Data points in the figure show the observational measurements of \taueffH reported by
\cite{Fan+2006}, \cite{Becker+2013a}, 
\cite{Bosman_2018}, \cite{eilers2018a}, \cite{boera2019a}, and \cite{Yang+2020b}.
Our results are consistent with the evolution of \HI\ \taueff measured by \cite{Becker+2013a} (yellow points) for the redshift range 
$2.5 \lesssim z \lesssim 4.2$.
Our model results in a more opaque IGM compared to their measurements at lower redshifts  $2.2 \lesssim z \lesssim 2.5$ and
higher redshifts
$ 4.2 \lesssim z \lesssim 4.8$.
Our model agrees well with the determination from \cite{boera2019a} (green points) during 
the redshift range $4.2 \lesssim z \lesssim 5.0$.
 
At high redshift ($z>5$), the measurements of the \HI\ effective optical depth from 
\cite{Bosman_2018} (red points), \cite{eilers2018a} (cyan points), and \cite{Fan+2006} (orange points)
are similar, with only small differences ($<12\%$) toward
higher \taueffH from \cite{eilers2018a} compared with \cite{Bosman_2018}.
The measurements by 
\cite{Yang+2020b} (purple points) suggest a more opaque IGM with a \taueffH
that is significantly higher ($\sim 20 - 30 \%$) compared to the measurements by \cite{Bosman_2018}. 

Shortly after hydrogen reionization completes ($5 \lesssim z \lesssim 5.8$),
our best-fit UVB model significantly underestimates \taueffH  compared with 
the observational measurements, suggesting that the hydrogen in the IGM is
overly ionized in our model at these redshifts.
To address this possible discrepancy, we can modify our best-fit result for the \HI\ 
photoionization rate such that $\Gamma_\mathrm{HI}$ is
reduced only in the redshift range $4.8 < z < 5.8$ and increased for $5.8 < z < 6.1$ (see \S \ref{sec:Gamma_HI } and \S \ref{sec:modified_uvb_model}). 
As shown in Figure \ref{fig:tau_HI}, the high redshift evolution ($z>5$) of  \taueffH from the modified model (dashed blue line) is consistent with the measurements from \cite{Bosman_2018}.
The subsequent evolution at redshifts $z<4.8$ remains virtually unchanged from the
best-fit model as hydrogen is in photoionization equilibrium at these times
and the ionization fraction is therefore determined
by the instantaneous amplitude of the \HI\ photoionization rate $\Gamma_\mathrm{HI}$.
We refer the reader to \S \ref{sec:modified_uvb_model} for a discussion on the effect that the
modified UVB model has on the properties of the gas in the IGM. 

By providing a simple modification to our best-fit UVB model that allows to change the high-redshift evolution of the hydrogen effective optical depth 
to achieve consistency with the observation and with minimal impact on the subsequent evolution of the properties of the IGM for $z\lesssim 5.0$, 
we show that the high-$z$ discrepancy of the observed \taueffH and the model is not a significant challenge to our results and the conclusions of this work.     

\begin{figure}
\includegraphics[width=0.47\textwidth]{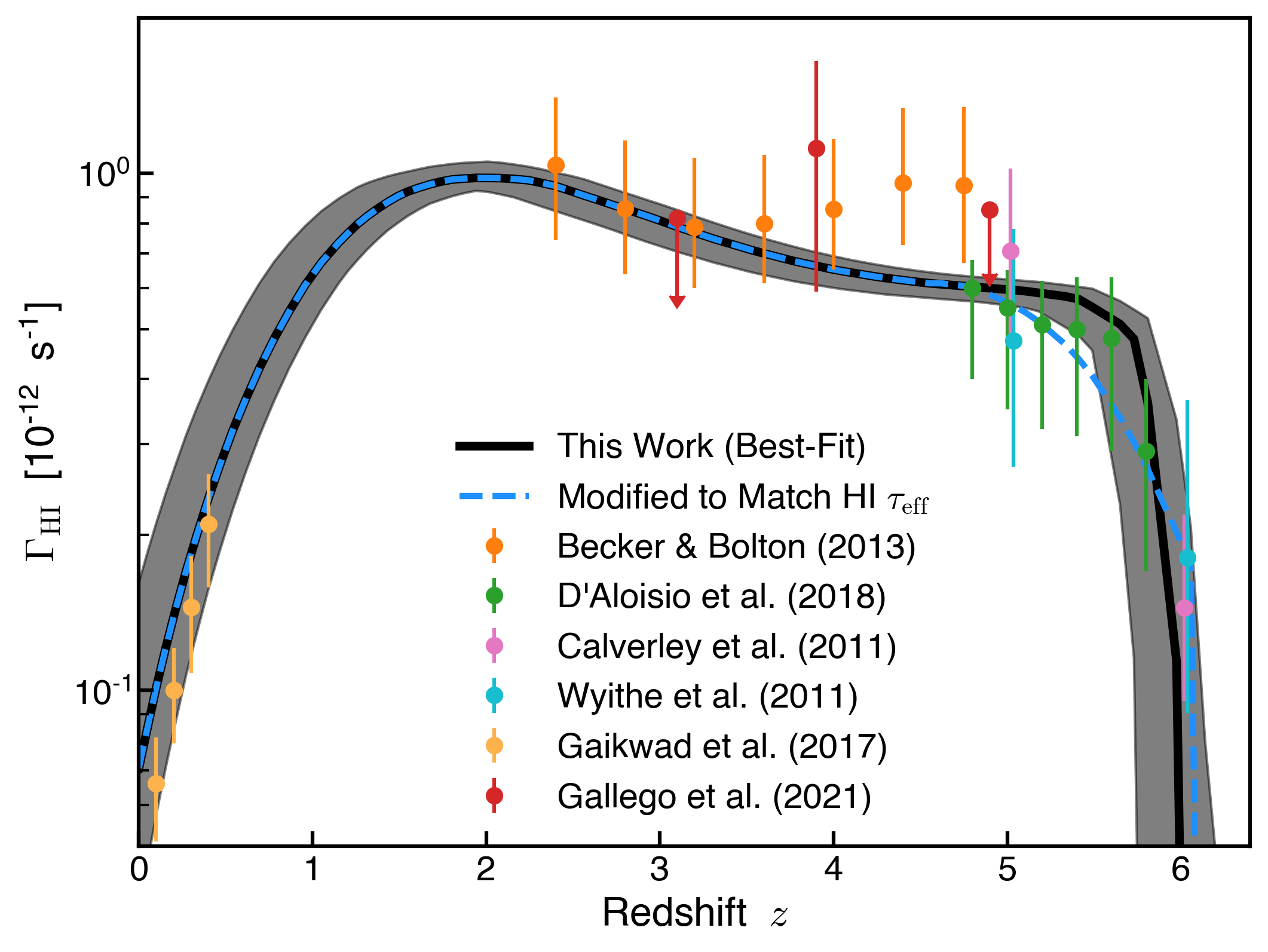}
\caption{ Evolution of the hydrogen photoionization rate \GammaHI from our best-fit determination and the 95\% confidence interval (black line and shaded region).
Data show observationally
inferred photoionization rates
measured by \cite{Calverley+2011}, \cite{Wyite+2011}, \cite{Becker+2013b}, \cite{Gaikwad+2017},\cite{daloisio2018a}, and \cite{Gallego+2021}. A modified model 
for \GammaHI designed to match the observational measurements of \taueffH from \citet[][see Figure \ref{fig:tau_HI}]{Bosman_2018} is shown as the dashed blue line.
Our models agree well with the observationally-inferred results,
except for visible differences
with the estimate from \cite{Becker+2013b} during $4 \lesssim z \lesssim 5$.
These differences in \GammaHI reflect
small differences between our best-fit model predictions for \taueffH and
the observational \taueffH measurement by \cite{Becker+2013a} over this redshift range.}
\label{fig:Gamma_HI}
\end{figure}

\begin{figure}
\includegraphics[width=0.47\textwidth]{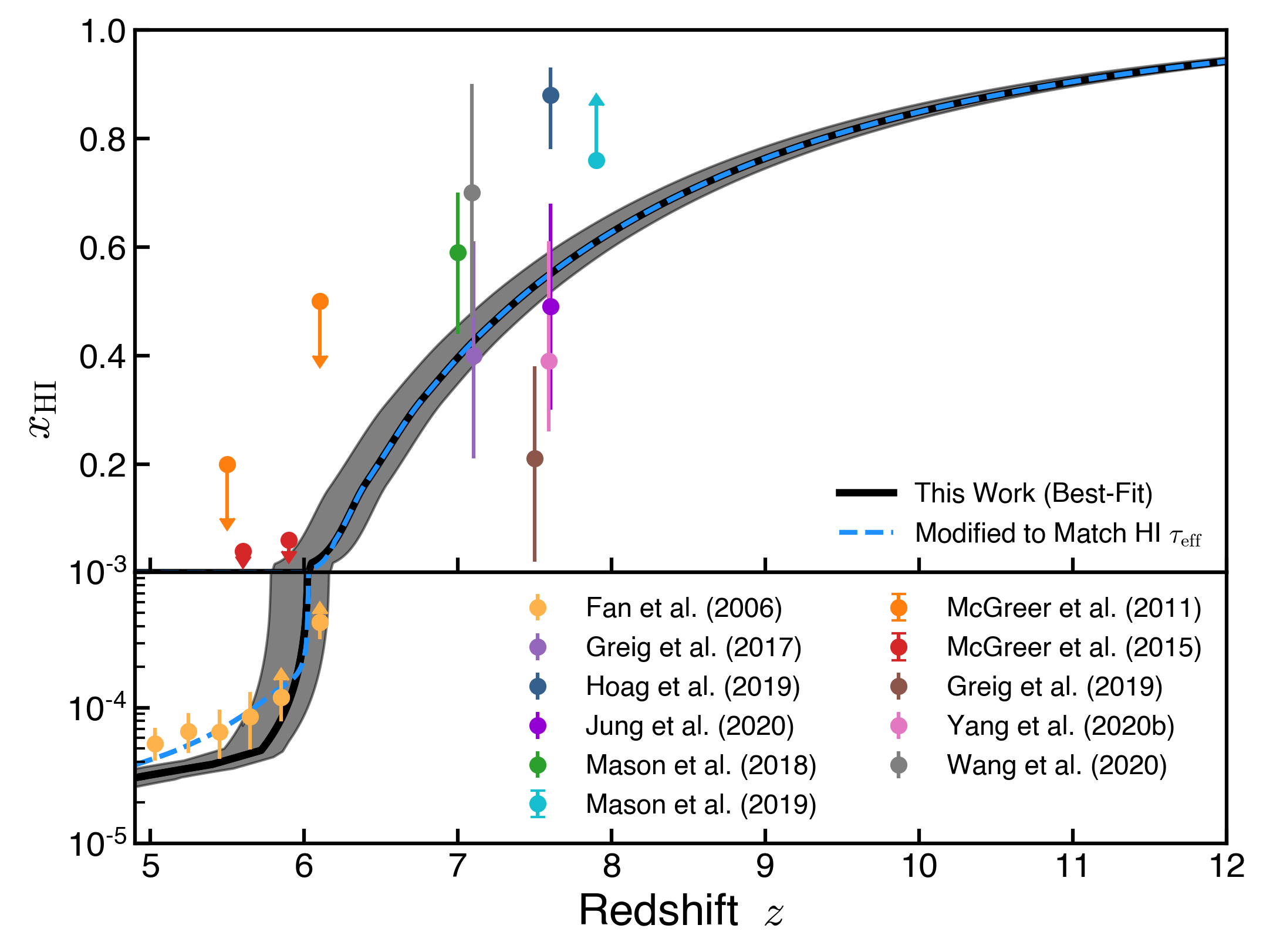}
\caption{Redshift evolution of the volume-weighted neutral fraction of hydrogen for our best-fit model and the corresponding 95\% confidence interval (black line and shaded region). 
Data points show the observational estimates reported in \cite{Fan+2006}, \cite{McGreer+2011, McGreer+2015}, \cite{Greig+2017, Greig+2019}, \cite{Mason+2018, Mason+2019},
 \cite{Hoag+2019}, \cite{jung2020}, \cite{Yang+2020a}, and \cite{Wang+2020}.
For $z\gtrsim 7$ the observational estimates show a wide range of \xHI, from \xHI$\sim 0.2$ to \xHI$\sim0.8$.
Our models result in a $z\sim7-8$ neutral fraction of \xHI$\sim0.4 -0.5$,
consistent with the results from \cite{Greig+2017}, \cite{jung2020}, and \cite{Yang+2020a}.
After hydrogen reionization completes at
$z\lesssim 6.0$,
our best-fit model shows an evolution of \xHI below the measurement by \cite{Fan+2006}.
By modifying our best-fit photoionization rates to better match 
\taueffH (see Figure \ref{fig:tau_HI}), we can also better match the
\xHI data from \cite{Fan+2006} (dashed blue line).}
\label{fig:HI_fraction}
\end{figure}

\subsection{Hydrogen Photoionization Rate}
\label{sec:Gamma_HI }

Our best-fit model results for the hydrogen photoionization rate \GammaHI provide
several opportunities for comparisons with observations, even though observationally
inferred \GammaHI measurements are not used to constrain our model.
There are observational determinations of \GammaHI
informed by simulations where the photoionization rate is rescaled to
match the observational \Fmean \citep{Becker+2013b, daloisio2018a}.
Our results can also be compared to estimates of \GammaHI from the quasar proximity effect and
the size of the near-zone of high \Lya\ transmission around quasars \citep{Calverley+2011, Wyite+2011}.
Observations have measured \GammaHI by detecting the florescent 
\Lya\ emission produced by the Lyman limit systems (LLS) illuminated by background radiation \citep{Gallego+2021}.
Finally, there are \GammaHI determinations from combining the PDF and power spectrum of the \Lya\ 
transmitted flux from observations with simulations that apply different photoionization rates
\GammaHI \citep{Gaikwad+2017}.

Figure \ref{fig:Gamma_HI} shows our result for the HI photoionization rate
with the corresponding 95\% confidence limits (black line and shaded band) along with the 
observational inferences of \GammaHI mentioned above.
Our result is consistent with the previous observational determinations
that show a rapid evolution in \GammaHI for $z \gtrsim 5.6$, followed by 
a gradual increase during $2 \lesssim z \lesssim 5.6$ and a rapid decrease at $z< 2$.
The only visible differences with \cite{Becker+2013b} occur in the 
redshift range $4 \lesssim z \lesssim 4.8$.
Their measurement was obtained by tuning the photoionization rate \GammaHI in simulations such that the Ly$\alpha$
effective optical 
depth \taueffH was consistent with the observational measurement from \cite{Becker+2013a}.
The higher estimate of \GammaHI from their result 
reflects the lower \taueffH from \cite{Becker+2013a} compared with the evolution of
\taueffH from our model for the redshift range $4.2 \lesssim z \lesssim 4.8$,
as shown in Figure \ref{fig:tau_HI}.

As described in \S \ref{sec:evolution_tau_HI }, shortly after hydrogen reionization 
completes 
our best-fit model
significantly underestimates
the \Lya\ effective optical depth \taueffH
compared with the observations in the redshift range $5 \lesssim z \lesssim 5.8$.
To address this discrepancy, we presented an alternative model where 
the sharp transition in \GammaHI at $z\sim 5.6$ from the original best-fit model
is replaced by a softer increase that extends over the redshift range $ 4.8 < z < 5.8$ (dashed 
blue line in Figure \ref{fig:Gamma_HI}). 
Decreasing \GammaHI during this epoch increases the neutral fraction of hydrogen in the IGM in photoionization equilibrium, thereby  increasing \taueff.
Our modified model for \GammaHI was chosen such that the resulting evolution of \taueffH is consistent with the observational measurement presented by 
\cite{Bosman_2018}(dashed blue line in Figure \ref{fig:tau_HI}), and the altered transition of \GammaHI from our modified model is still within the uncertainties of the observational inference 
by \cite{daloisio2018a} in the redshift interval $ 4.8 \lesssim z \lesssim 5.8$.

\begin{figure}
\includegraphics[width=0.47\textwidth]{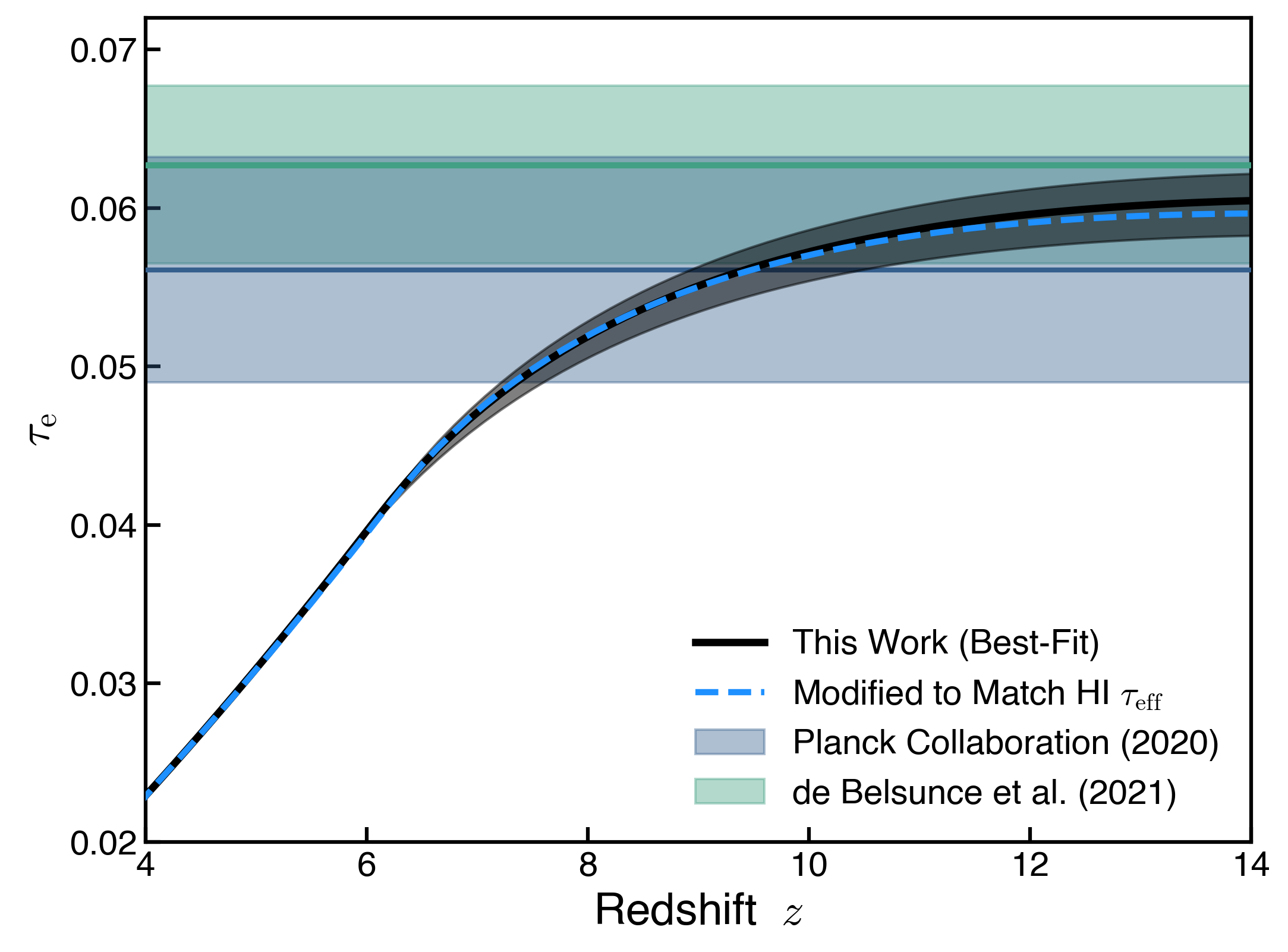}
\caption{Thomson optical depth from electron-scattering of the CMB $\tau_\mathrm{e}$
from the best-fit model and the 95\% confidence limit (black line and shaded bar) and our 
modified model to match the $z>5$ \taueffH (dashed blue line).
Also shown are the observational measurements from the Planck satellite presented in
\cite{Planck_collaboration_2020} 
and the constraint from \cite{deBelsunce+2021}.
Our model results for $\tau_\mathrm{e}$ lie within the Planck limits.}
\label{fig:tau_electron}
\end{figure}

\subsection{Ionization History}
\label{sec:ionization_history}

We present the redshift evolution of the
volume-weighted neutral fraction of hydrogen $x_\mathrm{HI}$ resulting from our best-fit determination of the UVB model and the 
corresponding 95\% confidence limits (black line and shaded band) in Figure \ref{fig:HI_fraction}.
For comparison we show several observational estimates. We show
constraints from
the optical depth of the Lyman-$\alpha$, Lyman-$\beta$, and  Lyman-$\gamma$ transitions in the forest \citep{Fan+2006}.
We also show constraints on the IGM neutrality from
properties of \Lya\ emission from galaxies 
at high redshift \citep{Hoag+2019, Mason+2018, Mason+2019} and
 the damping wing absorption 
 in the spectra of $z\gtrsim 7$ quasars \citep{Greig+2017, Greig+2019, jung2020, Yang+2020a,
Wang+2020}.
Finally, we show constraints from the
covering fraction of dark pixels in the Ly$\alpha/\beta$ forest of high-$z$ quasars \citep{McGreer+2011, McGreer+2015}. 

Our model results in a prolonged hydrogen reionization history, extending
from $x_\mathrm{HI} \sim 0.9  $ at $z\sim 11$ to $x_\mathrm{HI} \sim 0.1 $ at $z\sim 6.5$.
The duration results in part from
the gradually increasing ionization rate $\Gamma_\mathrm{HI} < 1\times 10^{-15}\, \mathrm{s^{-1}}$ at $z>6.5$
associated with radiation emitted by early star-forming galaxies. 
 
For $7 \lesssim z \lesssim 8$, the observational estimates display a wide range of $x_\mathrm{HI}$, from a highly ionized ($x_\mathrm{HI} \sim 0.8$) to a mostly neutral 
($x_\mathrm{HI} \sim 0.2$) IGM. Our model lies within this range,
and at $z=7$ our result is in agreement with the $x_\mathrm{HI}\sim 0.4$ estimates from \cite{Greig+2017} and
\cite{Yang+2020a} as well as with the $x_\mathrm{HI}\sim 0.5$ estimate from \cite{jung2020} at $z\sim 7.6$.

The redshift at which hydrogen reionization completes $z_\mathrm{R}$,
defined as the redshift at which $x_\mathrm{HI} \leq 1 \times 10^{-3}$ for the first time,
is $z\sim 6.0$ for 
our best-fit model. After hydrogen reionization completes,
our best-fit model results in an ionization fraction that
falls below the estimate from \cite{Fan+2006} (reflected by the lower 
optical depth \taueff in Figure \ref{fig:tau_HI}).
Nevertheless, our modified model (dashed blue line) shows better consistency with their estimate.

Later in  cosmic history, high energy radiation emitted by AGNs leads to the ionization of singly ionized helium 
(\HeII). For our best-fit model \HeII\ reionization starts at $z\sim5$ and completes at $z\sim 3.0$ when the \HeII\ fraction reaches
$x_\mathrm{HeII} \leq 1 \times 10^{-3}$ for the first time. As the \HeII\ effective optical depth from our model is consistent with the observation
from \cite{Worseck+2019} for $2.4 \lesssim z \lesssim 2.9$, we argue that the end of \HeII\ reionization by $z\sim 2.9$ is 
suggested by their measurement.   

Thomson scattering of the CMB by the free electrons in the IGM provides
another diagnostic of the reionization history of the IGM.
From the evolution of the electron density $n_\mathrm{e}$ given by the ionization state of hydrogen and helium from our models,
we can compute the electron scattering optical depth 
$\tau_\mathrm{e}$ as

\begin{equation}
\tau_{\mathrm{e}}(z)= \int_{0}^{z} \frac{c \sigma_{\mathrm{T}} n_\mathrm{e}(z)}{(1+z) H(z)} \mathrm{d} z
\label{eq:tau_electron}
\end{equation}         
\noindent
where $\sigma_\mathrm{T}$ represents the Thomson scattering cross section.
Figure \ref{fig:tau_electron} shows the
electron scattering optical depth $\tau_\mathrm{e}$ from our best-fit model
(black line and shaded region shows the 95\% confidence limit). Also shown are constraints
from the Planck satellite \citep{Planck_collaboration_2020} and the 
recent constraint from \cite{deBelsunce+2021}.
Our result for $\tau_\mathrm{e}=0.60$ lies
within the upper limit of the $\tau_\mathrm{e}=0.0540 \pm 0.0074$ constraint from
\cite{Planck_collaboration_2020} and in good agreement with the
determination of $\tau_\mathrm{e}=0.0627^{+0.0050}_{-0.0058}$ from \cite{deBelsunce+2021}.

\begin{figure}
\includegraphics[width=0.47\textwidth]{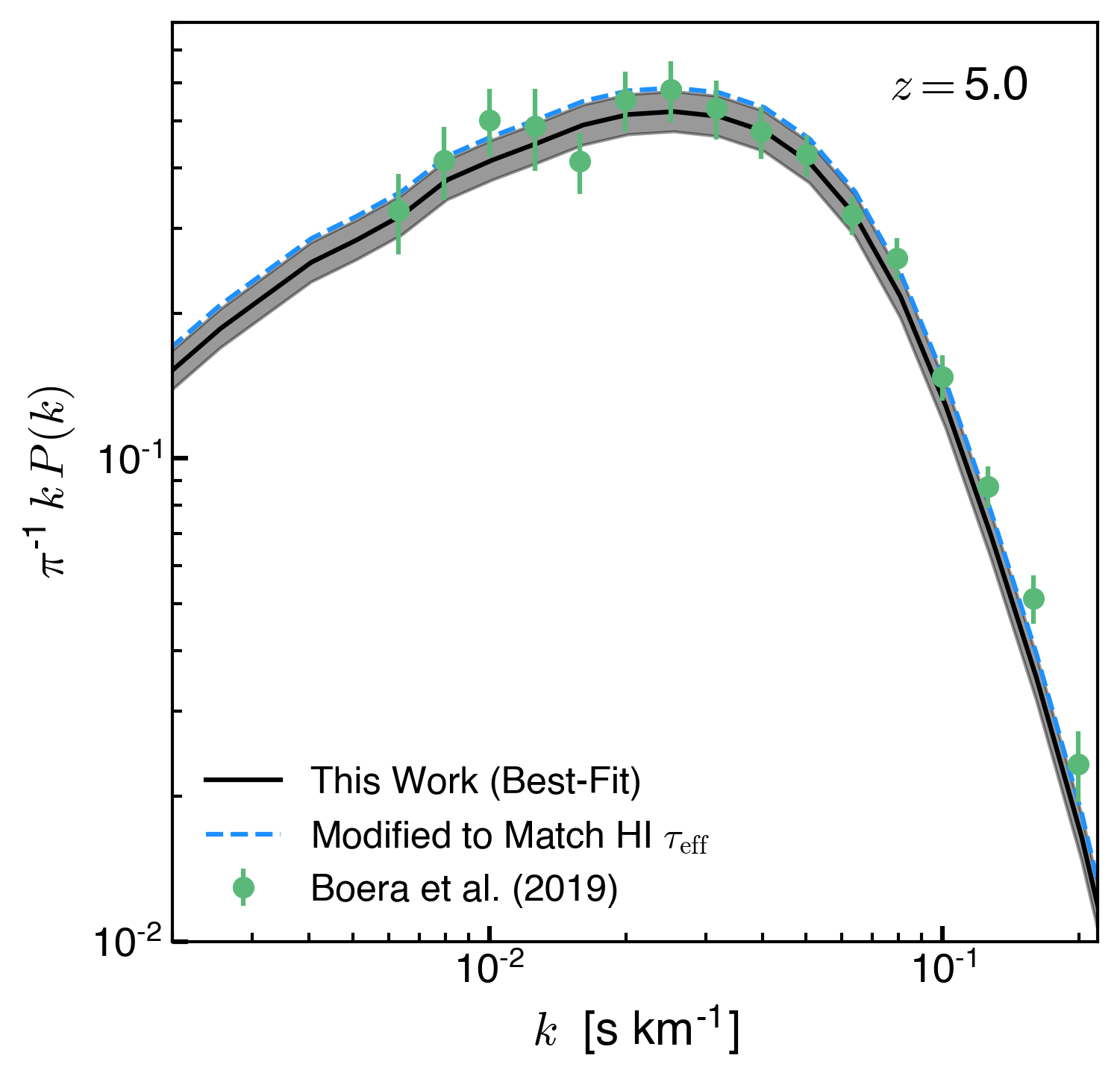}
\caption{Power spectrum of the \Lya\ transmitted flux $P(k)$ at $z=5$ from our best-fit model (black) and from our alternative model of the UVB (dashed blue) where the \HI\ and \HeI\
photoionization and photoheating rates are modified in the redshift range $4.8 \lesssim z \lesssim 6.1$ such that \taueffH is consistent with the observation from \cite{Bosman_2018}. 
The effect on the power spectrum from the modified model is to increase $P(k)$ by a roughly constant factor of $\sim 12\%$ compared with the best-fit model due to the $\sim 6\%$ 
increase in the \HI\ opacity at $z=5$. Both models are consistent with the observation from \cite{boera2019a} for $k \lesssim 0.1 \,\, \mathrm{s \, km^{-1}}$.}
\label{fig:power_spectrum_z5}
\end{figure}

\subsection{Modified UVB Rates for Matching the Observed High-Redshift Hydrogen Effective Optical Depth}
\label{sec:modified_uvb_model}

In \S \ref{sec:evolution_tau_HI } and \S \ref{sec:Gamma_HI } we discuss how 
the IGM from our best-fit model is possibly
too highly ionized after hydrogen reionization completes.
The hydrogen effective optical depth \taueffH from the model is
significantly lower compared with observations in the redshift range $5 \lesssim z \lesssim 5.8$ (see Figure \ref{fig:tau_HI}).
We can 
address this issue by decreasing the \HI\ photoionization rate \GammaHI such that the sharp transition
at $z\sim 5.8$ from the best-fit 
model is replaced by a more gradual increase of \GammaHI during the redshift range
$4.8 \lesssim z \lesssim 6.0$ (dashed blue line in Figure \ref{fig:Gamma_HI}). 
This alternative transition in \GammaHI was chosen such that
the resulting evolution of \HI\ \taueffH is consistent with the observations
from \cite{Bosman_2018}.

Assuming that changes made to the photoionization rate \GammaHI correspond to a change of the mean-free-path 
of ionizing photons $\lambda_\mathrm{mfp}$, then the \HeI photoionization rate $\Gamma_\mathrm{HeI}$ 
should also reflect the modification applied to \GammaHI.
Correspondingly, we rescale the helium photoionization rate $\Gamma_\mathrm{HeI}$ such that
the ratio
$\Gamma_\mathrm{HI}(z) / \Gamma_\mathrm{HeI}(z) $ from the modified model matches the best-fit model.

Changing $\lambda_\mathrm{mfp}$ would also affect the photoheating rates
$\mathcal{H}_\mathrm{HI}$ and $\mathcal{H}_\mathrm{HeI}$.
Assuming 
that the average energy of the ionizing photons remains the same in the modified model,
we rescale the photoheating rates such that the  
ratios $\mathcal{H}_\mathrm{HI}(z) / \Gamma_\mathrm{HI}(z) $  and $\mathcal{H}_\mathrm{HeI}(z) / \Gamma_\mathrm{HeI}(z)$
match the best-fit model.
Results from our modified model for photoheating and photoionization rates are shown in
Figure \ref{fig:uvb_result} as dashed blue lines.

After hydrogen reionization completes at
$z \lesssim 6.0$,
hydrogen in the IGM is in photoionization equilibrium.
During this epoch, decreasing the \HI\ and \HeI\ photoionization 
rates effectively increases the the neutral fraction of hydrogen and helium.
Consequently the opacity of the IGM, quantified as the optical depth \taueff, also increases during 
redshift range.
The temperature of the gas in the IGM is not strongly affected by the modified photoionization and photoheating rates,
because, in equilibrium, the 
gas temperature $T(z) \propto \mathcal{H}(z) / \Gamma(z)$ and this ratio is unchanged
from the best-fit model. 

The modified model only changes the photoionization and photoheating rates during the redshift range $4.8 \leq z \leq 6.1$.
These changes result
in an increase of \taueffH during $4.8 \leq z \leq 5.8$ and a decrease during $5.8 < z \leq 6.1$ but do not strongly affect
the evolution of the gas temperature. For redshifts $ z < 4.8$,
the ionization fraction of 
hydrogen in the IGM in photoionization equilibrium is determined by the ratio of the photoionization rate to the recombination rate 
$x_\mathrm{HII}(z) \propto \Gamma_\mathrm{HI}(z) / \alpha_\mathrm{HII}(z,T)$. 
Since the thermal evolution resulting from the modified 
and best-fit models are very similar and the rates $\Gamma$ and $\mathcal{H}$ at  $ z < 4.8$ are the same. Thereby, 
the evolution of the neutral fraction $x_\mathrm{HI}$,
the effective optical depth \taueffH, and the \Lya\ power spectrum $P(k)$ resulting from the 
modified model is nearly unchanged from the best-fit model at redshifts $z < 4.8$.                   

The increase in the hydrogen effective optical depth \taueffH during 
the redshift range $4.8 \leq z \leq 5.8$ in the modified model influences the \Lya\ 
power spectrum at this epoch.
Given the available data, this modification
only affects comparisons with the observed $P(k)$ at $z=5.0$.
Figure
\ref{fig:power_spectrum_z5} shows $P(k)$ from the modified (dashed blue) and best-fit model (black) at $z=5$.
Relative to the best-fit model, using the modified model results in a small
increase ($\sim 12\%$) in $P(k)$ owing to the small increase ($\sim 6\%$) in \taueffH. 
Either model shows consistency with the observational $P(k)$ measurement from \cite{boera2019a}.

\subsection{Limitations of the Model }\label{sec:limitations}

For this work, we have modeled the evolution of the properties of the IGM using a spatially homogeneous ionizing
background. Simulations of a more realistic, spatially inhomogeneous hydrogen reionization
process show that spatial fluctuations in the temperature–density relation of the post-reionization
IGM have a minor effect on the flux power spectrum \citep{Keating+2018} at $z\leq 5$ 
while the inhomogeneous UVB allows large islands of neutral hydrogen to persist up to redshift $z \leq 5.5$ and can reproduce 
the observed distribution of \Lya\ opacity \citep{kulkarni2019}. 
Similarly, radiative transfers simulations of
\HeII\ reionization show that the fluctuations in the ionization state of helium have a minor effect on observations of the hydrogen 
\Lya\ forest \citep{laplante2017a, Upton-Sanderbeck+2020}. Not including the impact of galactic winds or AGN-feedback on the forest 
is a conservative approach for simulations aimed at constraining effects that suppress small-scale power. AGN feedback in the form 
of heating or mass redistribution from small to large scales is also expected to suppress the 1D power spectrum on large scales, 
and to have an increased effect at low redshifts \citep{Viel+2013}. Ignoring the impact of AGN feedback may lead to a few 
percent bias in the determination of cosmological and astrophysical parameters \citep{Chabanier+2020}. This model uncertainty is 
comparable to the statistical uncertainties of the eBOSS data used in this work.

Another limitation of our method results from the UVB photoionization and photoheating rates
used for our simulation grid being constructed from simple transformations of a template set of rates.
We therefore do not probe the full 
range of ionization and thermal histories that could be allowed by the observations of 
the \Lya\ forest.
However, our model produces statistical properties 
of the \Lya\ forest that agree 
with a wide range of observations and a thermal evolution of the IGM consistent with previous inferences.
These features of our work represent
a significant achievement enabled by the ability to
explore a wide range of models for the UVB from self-consistently evolved simulations.
We emphasize that with our computational capabilities,
performing a very large number of simulations (e.g., thousands) is now a possibility. We therefore defer more 
flexible explorations of models for the heating and ionization from the UVB to future work.  

In the approach used for this work, we modify the photoionization and photoheating jointly.
This joint variation results in another important limitation of our study. The large scales of the power spectrum of the 
forest are sensitive to the ionization state of \HI\ which, in equilibrium, is set by the balance between 
photoionization and recombination. The large-scales of $P(k)$ depend on the temperature of the gas
through the recombination coefficient $\alpha(T) \propto T^{-0.72}$ but are mostly determined by the intensity of the
photoionization rate $\Gamma_\mathrm{HI}$. Since a large fraction of the dataset used for our inference probes the
large-scale $P(k)$ the best-fit photoheating rates are influenced by the determination of the best-fit photoionization 
rates. We have shown that the photoheating from our best-fit model is consistent with other estimates of the 
thermal state of the IGM determined independently.
Nevertheless, the relatively small uncertainty in the thermal state parameters $T_0$ and $\gamma$ from this work is in part 
a consequence of the
of the well-constrained determination of the photoionization rate from the large-scale $P(k)$. In future work we will explore 
a more flexible approach in which the photoheating has some degree of freedom with respect to the photoionization rate, such
as using density-dependent UVB rates to better model a inhomogeneous reionization.

\section{Summary}
\label{sec:summary}

With the objective of finding a photoionization and photoheating history that results in properties of the IGM consistent with
observations of the hydrogen and helium  \Lya\ forest,
we have used the  GPU-native \Cholla code to perform an unprecedented grid of more than 400 cosmological simulations
spanning a variety of ionization and thermal histories of the IGM. 
These calculations extend our \Sim suite of hydrodynamical simulations initially presented 
in \cite{villasenor2021a}. 
We compare the properties of the \Lya\ forest from our 
simulations to several observational measurements to determine via a likelihood
analysis the best-fit model for the photoionization and photoheating rates.
From our best-fit model we have inferred the thermal history of the IGM, and
demonstrate consistency with recent estimates obtained from the properties of the \Lya\ forest.
A summary of the efforts and conclusions from this work follows.

\begin{itemize}

\item We present a direct extension of the \Sim suite \citep{villasenor2021a} consisting of a grid of 400 simulations 
($L=50\,h^{-1}$Mpc, $N=1024^3$) that vary the spatially-uniform photoionization and photoheating rates from 
the metagalactic UVB.  The UVB rates applied for our grid use the \cite{puchwein2019a} model as a template, and
use four parameters that control a rescaling of the amplitude and redshift-timing of the hydrogen and helium
photoionization and photoheating rates.

\item The \Sim simulations self-consistently evolve a wide range of ionization and thermal histories of the IGM. 
We compare the properties of the \Lya\ forest in the form of the power spectrum $P(k)$ of the hydrogen \Lya\ transmitted 
flux and the helium (\HeII) effective optical depth \taueffHe from our simulations to several observational measurements covering 
the redshift range $2.2 \leq z \leq 5.0$ for $P(k)$ \citep{Irsic+2017a, boera2019a, Chabanier+2019} and $2.4 \lesssim z \lesssim 2.9$ for  \taueffHe  \citep{worseck2016a}.    

\item We perform a Bayesian MCMC marginalization to determine the best-fit UVB model. The performance of each  model in reproducing the
observations is evaluated over the entire redshift evolution instead of comparing for each redshift bin independently. Additionally, 
our simulation grid naturally probes a large range of ionization histories that we match directly to evolution of the ionization 
state of hydrogen encoded in the power spectrum of the \Lya\ forest. We thereby avoid any need to rescale the optical depth from the 
simulations in post-processing to match the observed mean transmission of the forest, which is a common shortcoming of previous 
analyses.

\item Our approach does not require an assumption of a power-law relation for the density-temperature distribution of the gas, as the \Lya\ spectra
is constructed from our self-consistently evolved simulations. We
find that a single power law does not accurately describe the $\rho_\mathrm{gas}-T$ 
distribution of the gas in the density range relevant to generating the signal of the \Lya\ forest. 

\item From our analysis, we infer the evolution of the thermal state of the IGM. The temperature
history of the IGM shows a first temperature peak ($T_0 \simeq 1.3 \times 10^4 \mathrm{K}$) due to
hydrogen reionization at $z\simeq 6$. This peak is followed by an epoch of cooling due to adiabatic
expansion of the Universe until the onset of helium reionization 
from radiation emitted by AGNs. The ionization of helium leads to a second increase of the
temperature until \HeII\ is fully ionized ($z\simeq 3$), resulting 
in a second peak of $T_0\simeq 1.4 \times 10^4 \mathrm{K}$. The second peak is
followed by a second period of cooling from cosmic expansion.    
Our result is consistent with previous estimates from \cite{gaikwad2020a} and \cite{gaikwad2020b}.
We note that the method employed in this work where we modify the UVB photoionization and photoheating rates
by rescaling and shifting the model from \cite{puchwein2019a} limits the variation on the evolution of 
the thermal history of the IGM in our simulations. In future work we will allow for more flexibility in the photoheating history
which will result in a more complete sample of the IGM density-temperature distribution. The improved flexibility of the models may permit a better inference of the thermal history of the IGM, as for now our low-redshift ($z < 4$) constraints are largely informed by the 
ionization state of hydrogen which likely results in a underestimated uncertainty in our $T_0 - \gamma$ evolution.

\item We compare the evolution of the hydrogen effective optical depth \taueffH
from our best-fit model to several observational determinations.
We find that after hydrogen reionization completes
($5 \lesssim z \lesssim 6$), the \HI\ effective optical depth resulting from the model may
underestimate the observations. We provide a modification to our best-fit model where the photoionization and photoheating 
rates are reduced during this epoch such that the evolution of \taueffH is consistent with measurements by \cite{Bosman_2018}. 
Additionally, the neutral fraction of hydrogen from the modified model shows consistency with the measurements
by \cite{Fan+2006} during this redshift interval.

\item The model for the photoionization and photoheating rates from the UVB obtained from our analysis shows consistency
with the observations of the \Lya\  
power spectrum and the effective optical depth from both hydrogen and helium (\HeII), the optical depth from the
CMB probed by \cite{Planck_collaboration_2020},
and previous inferences of the thermal state of the IGM.
This model can be applied in future cosmological simulations that aim to reproduce properties of the IGM
consistent with the observed \Lya\ forest.

\end{itemize}

Our work shows that an exploration of the IGM properties from hundreds of self-consistently evolved models for the astrophysical processes
that impact the gas in the medium is now possible by exploiting modern computational techniques on the world's largest supercomputers.
Using our efficient GPU-based code \Cholla with {\it Summit},
we are able to run hundreds of cosmological simulations in just a few days using a small fraction of the system.
We anticipate that when combined 
with the exquisite picture of the \Lya\ forest that experiments like \cite{desi2016a} will provide,
this capability will revolutionize future studies of the properties of the 
IGM.
We can leverage next-generation exascale systems and simulate large volumes ($L\sim 50h{-1}$Mpc) at high resolution ($N=2048^3$) 
for thousands of models describing the various the astrophysical processes that affect the IGM with
 a range of cosmological parameters, and study 
different models for the nature of dark matter and the mass hierarchy of neutrinos
based on their impact on the small-scale power spectrum of the \Lya\ forest.

\smallskip
\smallskip

\textit{Software:} \Cholla \citep[\url{https://github.com/cholla-hydro/cholla}]{schneider2015a}, Python \citep{Python}, Numpy \citep{numpy}, 
Matplotlib \citep{matplotlib}, MUSIC \citep{Hahn+2011_Music}, GRACKLE \citep{smith2017a}.

\acknowledgements
This research used resources of the Oak Ridge Leadership Computing Facility at the Oak Ridge National Laboratory, which is supported by 
the Office of Science of the U.S. Department of Energy under Contract DE-AC05-00OR22725, using Summit allocations CSC434 and 
AST169.  An award of computer time was provided by the INCITE program, via project AST175. We acknowledge use of the {\it lux} supercomputer at UC Santa Cruz, funded by NSF MRI grant AST1828315, and support from 
NASA TCAN grant 80NSSC21K0271. B.V. is supported in part by the UC MEXUS-CONACyT doctoral fellowship. B.E.R. acknowledges 
support from NASA contract NNG16PJ25C and grants 80NSSC18K0563 and 80NSSC22K0814.  We acknowledge the comments
and suggestions received from the anonymous referee which helped improve the content and clarity of this
work.

\appendix

\begin{figure*}
\includegraphics[width=\textwidth]{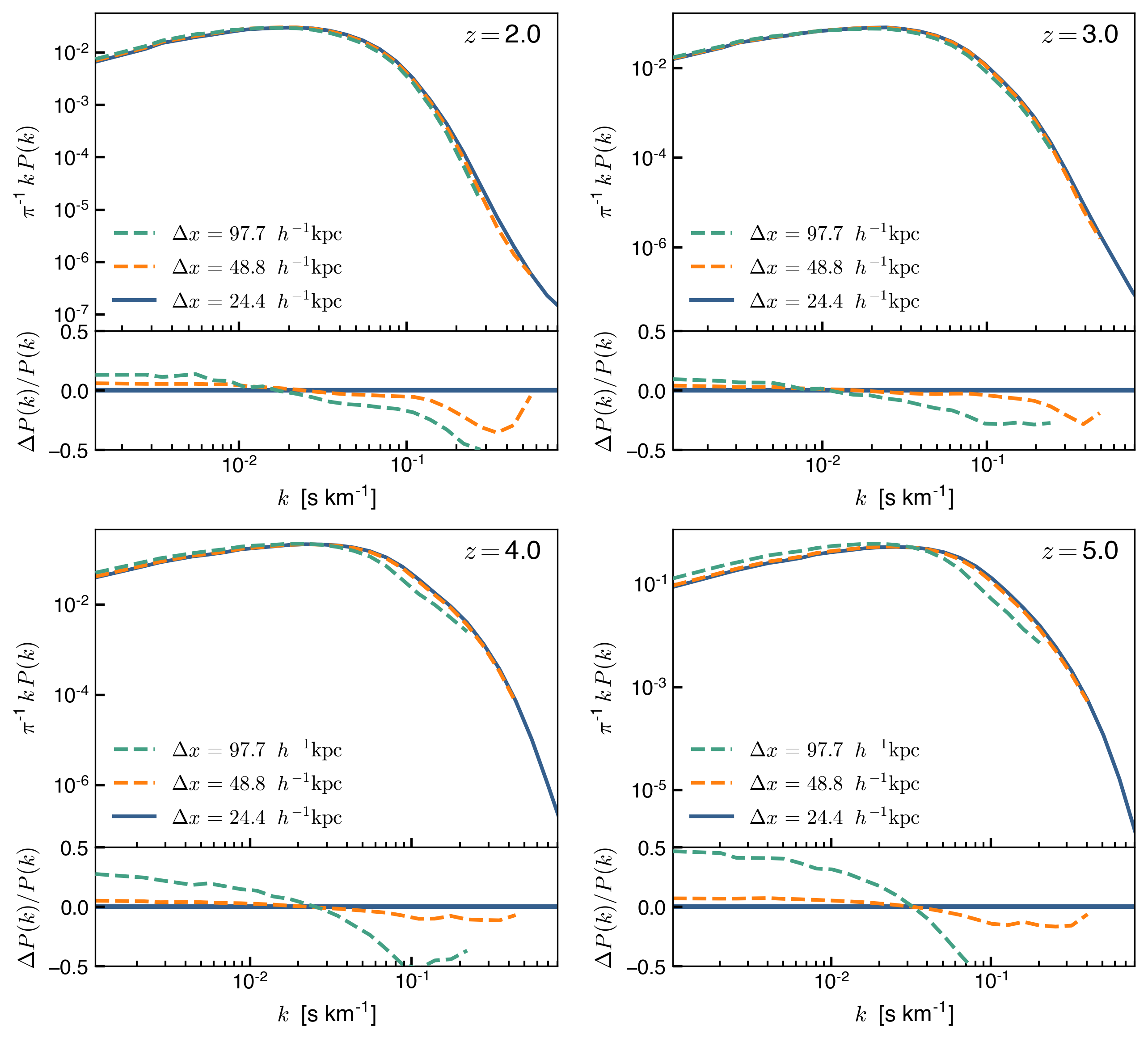}
\caption{Power spectrum of the \Lya\ transmitted flux $P(k)$ measured from simulations
with different comoving spatial resolutions of 
$\Delta x\,\simeq$ 98, 49, and 24 $h^{-1}$kpc.
The three simulations model a $L=50h{-1}$Mpc box with the \cite{Planck_collaboration_2020} 
cosmology and apply our best-fit determination for
the photoionization and photoheating rates.
The bottom panels show the fractional difference 
in the power spectrum  $\Delta P(k) / P(k)$ between the $N=512^3$ and $N=1024^3$
runs and the $N=2048^3$ simulation. 
Low-resolution simulations show increased power on
large scales ($k \lesssim 0.03 \,\, \mathrm{ s\, km^{-1}}$) and suppressed
structure in the small scales relative to higher resolution simulations. 
For the intermediate-resolution simulation
$N=1024^3$, which corresponds to our fiducial \Sim grid resolution,
the differences in $P(k)$ with respect to
the $N=2048^3$
simulation are $\lesssim 7\%$ on the large scales 
and $\lesssim 10-25\%$ on the small scales. We account for this resolution effect
during our inference procedure by adding a systematic error 
to the observational measurements of $P(k)$ in the form of
$\sigma_\mathrm{res}= \Delta P(k,z)$, where $\Delta P(k,z)$ is the redshift- and scale-dependent 
difference in the power spectrum measured from the $N=1024^3$ run
compared with the $N=2048^3$ simulation.}  
\label{fig:resolution}
\end{figure*}

\section{Resolution Convergence Analysis}
\label{sec:resolution}

To assess the possible impact of the simulation spatial resolution on our results,
we compare the \Lya\ transmitted flux power 
spectrum measured from simulations with different resolutions.
Each run was performed using the same box size ($L=50 h^{-1}\mathrm{cMpc}$) 
for identical cosmological parameters \citep{Planck_collaboration_2020} and our best-fit determination for the photoionization and photoheating rates,
and differ only in their grid resolution.
Our comparison is made between three runs with resolutions $N=512^3$, $N=1024^3$, 
and $N=2048^3$ cells and dark matter particles, with comoving spatial resolutions
of $\Delta x\simeq$ 98, 49, and 24 $h^{-1}$kpc, respectively. 
The initial conditions for the runs were generated to preserve common large-scale modes,
such that the results from the 
simulations could be compared directly over shared spatial scales.

Figure \ref{fig:resolution} shows the power spectrum of the \Lya\ flux measured
for our three simulations at redshifts $z=$2, 3, 4, and 5. As shown, the structure 
of the \Lya\ forest becomes better resolved as the number of cells  increase.
The lower panels present the fractional difference $\Delta P(k) /P(k)$ of the power 
spectrum measured from the $N=512^3$ and $N=1024^3$ simulations compared with
the $N=2048^3$ simulation on  overlapping spatial scales. Our comparison 
shows that the effect of the decreased resolution is to increase the power on large scales ($k \lesssim 0.02 \,\, \mathrm{s\, km^{-1}}$) while the small-scale
power is suppressed. For the low-resolution simulation ($N=512^3$) the differences are significant, and
on large scales the power spectrum is overestimated 
by $\sim 50\%$ at redshift $z=5$.
As the redshift decreases
the differences also decrease to $\sim 13\%$ by $z=2$. On small scales, the power 
spectrum is suppressed by $20-60\%$.

Our fiducial resolution for the \Sim simulations was
$N=1024^3$. At this resolution we measure only small
differences in the \Lya\ structure compared with the $N=2048^3$ simulation,
as on large spatial scales the power spectrum is overestimated by $\lesssim 7\%$,
and for small scales ($0.03 \lesssim k \lesssim 0.2 \, \, \mathrm{s \, km^{-1}}$) 
we measure a suppression on $P(k)$ of $\lesssim 10- 25\%$. To account for the effect of
resolution on simulations used to constrain the 
UVB model, we include a systematic uncertainty of
the form $\sigma_\mathrm{res}= \Delta P(k,z)$, where $\Delta P(k,z)$ is the redshift- and
scale-dependent difference in the power 
spectrum measured between the $N=1024^3$ and $N=2048^3$ simulations.

\begin{figure*}
\includegraphics[width=\textwidth]{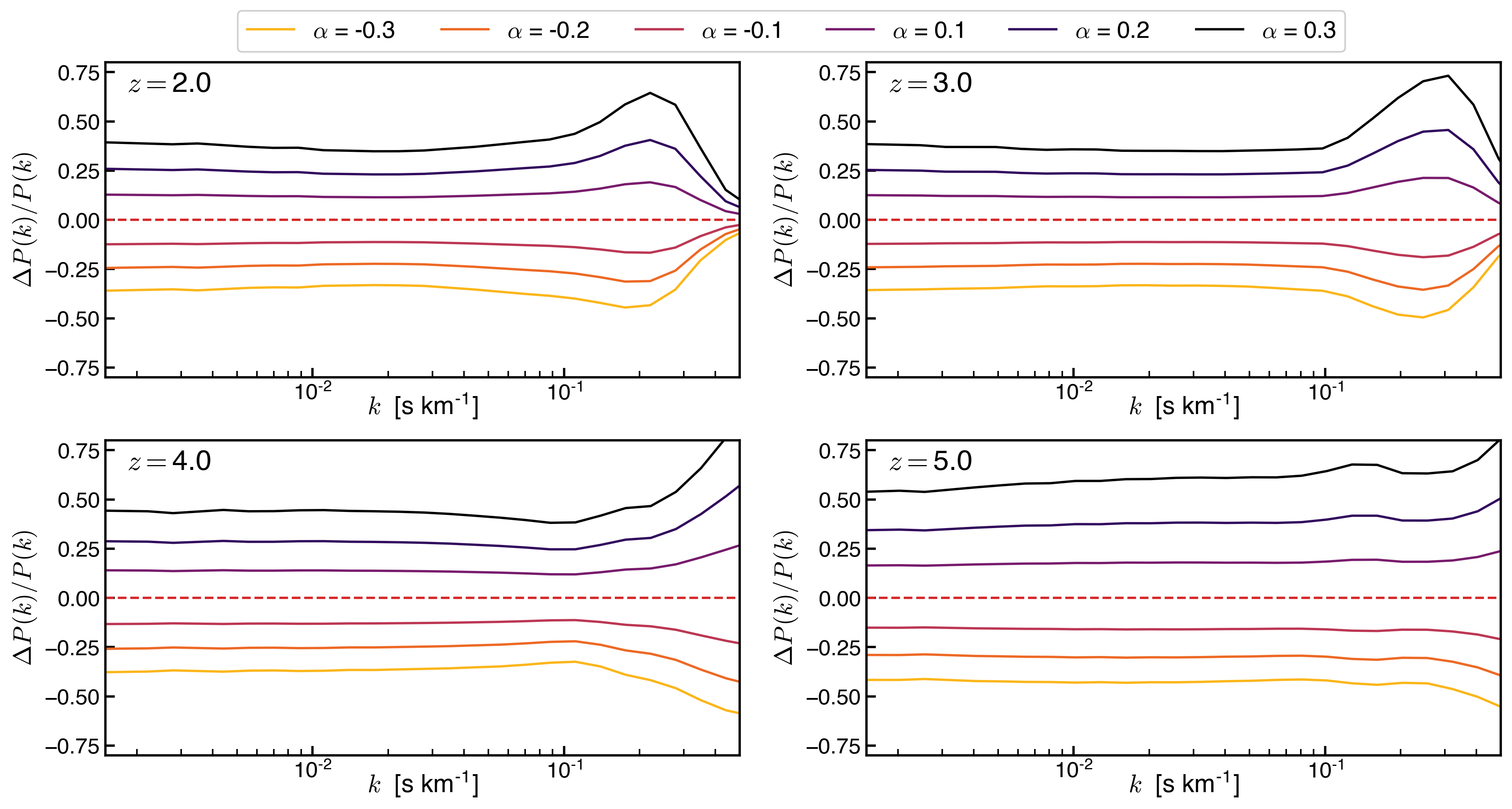}
\caption{Consequences of rescaling the effective optical depth
for the power spectrum of the \Lya\
transmitted flux at redshifts
$z=$2, 3, 4, and 5. Shown is
the fractional difference $\Delta P(k) / P(k)$ after rescaling the optical 
depth along the skewer sample from our simulations 
by a constant factor such that $\widetilde{\tau_\mathrm{eff,H}} = ( 1 + \alpha ) \tau_\mathrm{eff}$ 
for $\alpha$ in the range [-3, 3].
Rescaling the optical depth along the skewers 
such that \taueffH increases (decreases) has the effect of increasing (decreasing) $P(k)$.
On scales in the range $0.002 \lesssim k \lesssim 0.1 \,\, \mathrm{s \, km^{-1}}$
the change induced on $P(k)$ is almost uniform, while for the smallest scales
$k \gtrsim 0.1 \,\, \mathrm{s \, km^{-1}}$ the effect is redshift- and scale-dependent.}
\label{fig:rescaled_power_spectrum}
\end{figure*}

\begin{figure*}
\includegraphics[width=\textwidth]{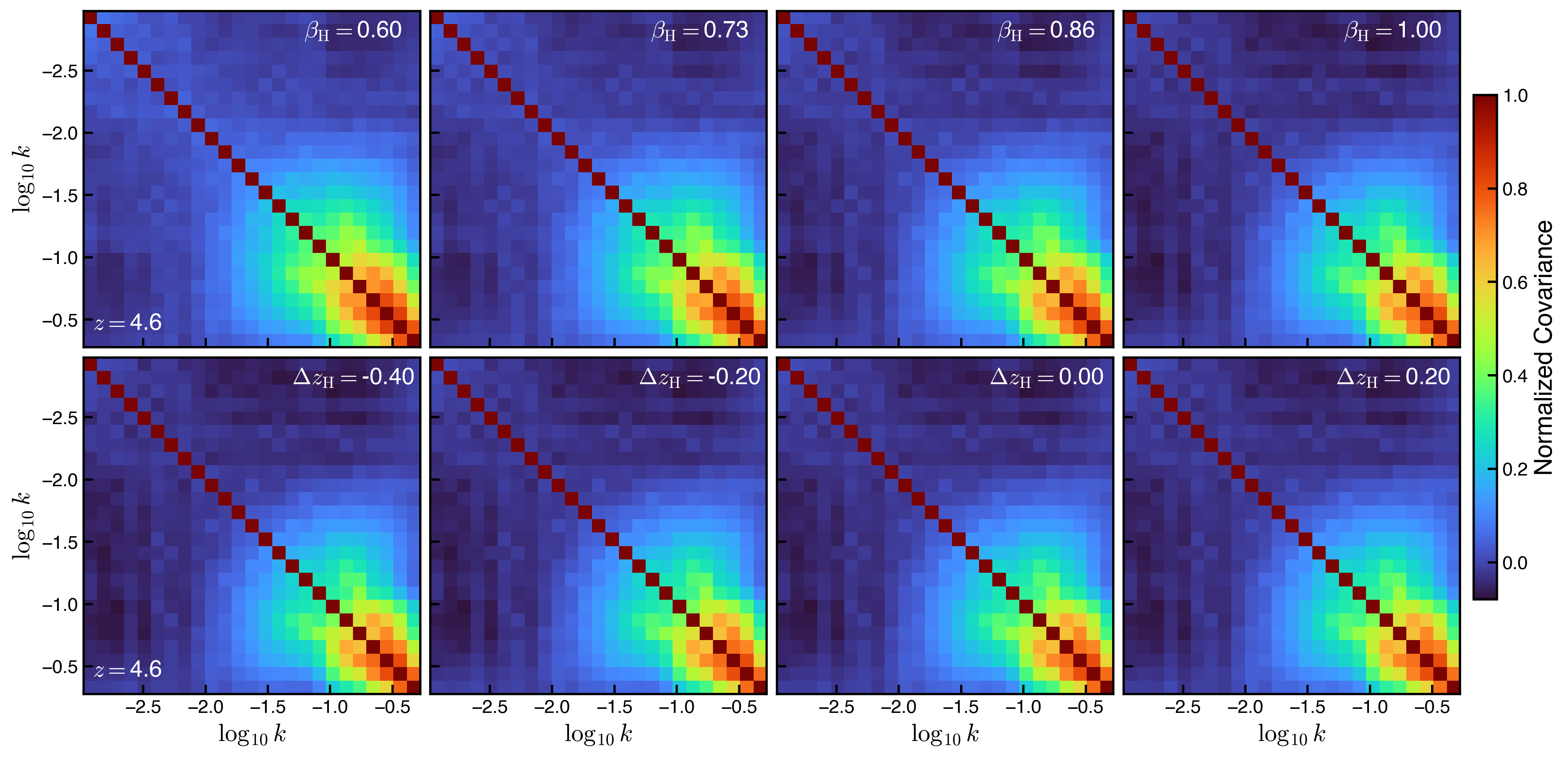}
\caption{
Normalized covariance matrix of the \Lya\ transmitted flux power spectrum at $z=4.6$ measured from simulations that 
vary the parameters $\beta_{\mathrm{H}}$ (top panels) and $\Delta z_{\mathrm{H}}$ (bottom panels) independently. 
The structure of the covariance is maintained across the simulations.
Decreasing the parameter $\beta_{\mathrm{H}}$ increases the normalization of $P(k)$ and its covariance on 
roughly all scales.  We measure small elementwise differences $<0.1$ in the normalized covariance across 
simulations with different $\beta_{\mathrm{H}}$. The effect of
changing $\Delta z_{\mathrm{H}}$ is minimal with elementwise differences $<0.03$. }
\label{fig:covariance_matrix_H}
\end{figure*}

\begin{figure*}
\includegraphics[width=\textwidth]{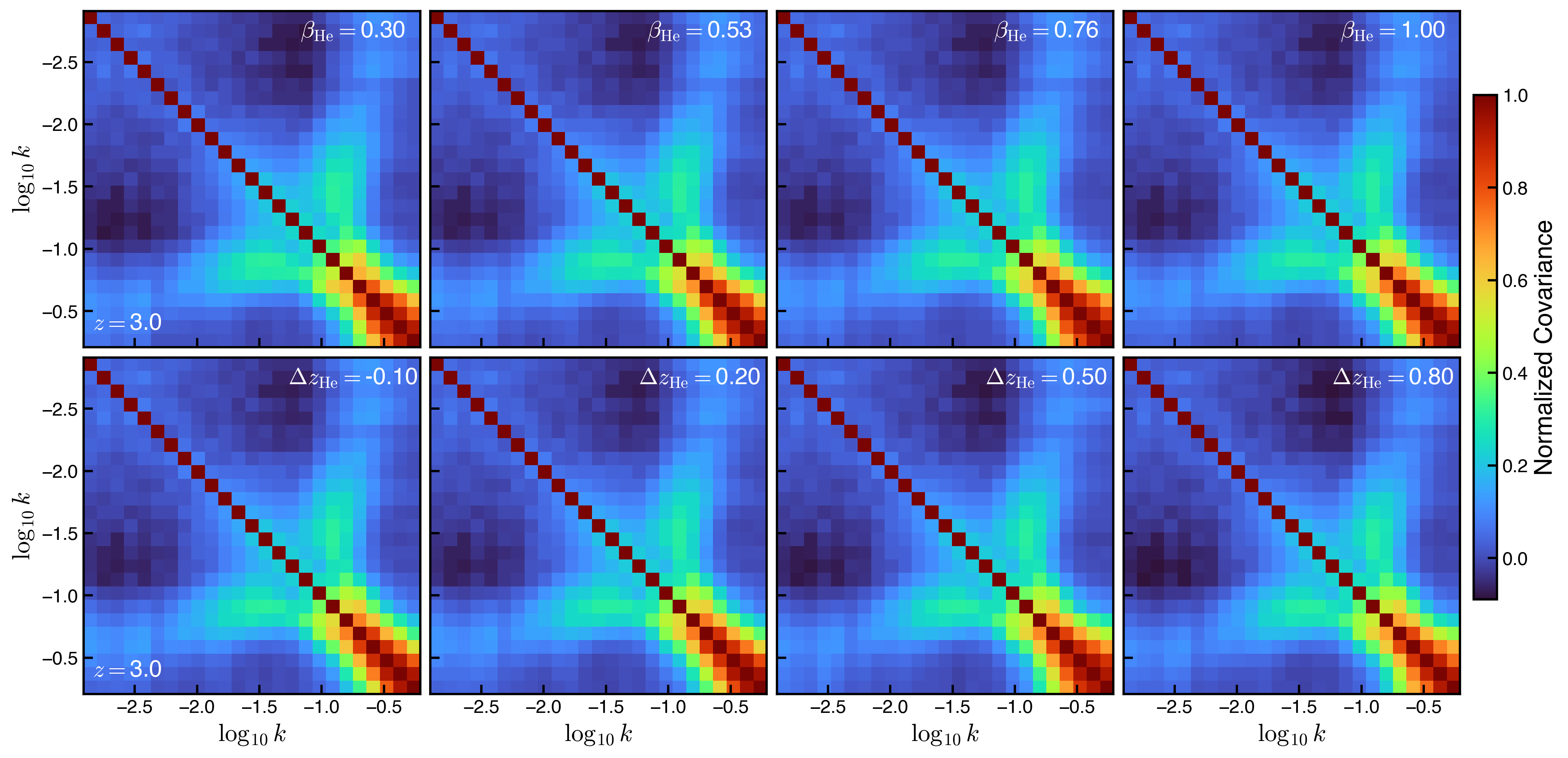}
\caption{
Normalized covariance matrix of the \Lya\ transmitted flux power spectrum at $z=3.0$ measured from simulations that 
vary the parameters $\beta_{\mathrm{He}}$ (top panels) and $\Delta z_{\mathrm{He}}$ (bottom panels) independently. 
The structure of the covariance is maintained across the simulations. Changes in $\beta_{\mathrm{He}}$ and $\Delta z_{\mathrm{He}}$
cause small variation in the normalized covariance matrix, we measure only small elementwise differences $<0.05$ over these simulations.   }
\label{fig:covariance_matrix_He}
\end{figure*}

\section{Effect of Rescaling the \HI\ Effective Optical Depth on the Lyman-$\alpha$ Flux Power Spectrum}
\label{sec:rescale_tau}

The power spectrum of the \Lya\ transmitted flux $P(k)$ is computed from
flux fluctuations $\delta_F = ( F - \langle F \rangle) / \langle F \rangle$.
The power spectrum is sensitive to changes on the ionization state of
hydrogen in the IGM, which in turn changes the effective optical 
depth \taueffH and the mean transmitted flux $\langle F \rangle = \exp(- \tau_\mathrm{eff}) $.
To estimate how changes in the overall ionization state of the IGM
affect the power spectrum of the \Lya\ flux, we can rescale the
optical depth of the simulated skewers and re-measure $P(k)$.
We rescale by a constant factor 
tuned such that the effective optical depth measured from the rescaled skewers follows 
$\widetilde{\tau_\mathrm{eff,H}} = ( 1 + \alpha ) \tau_\mathrm{eff,H}$, where \taueffH is the original effective optical depth obtained from the simulated skewers. 
From the rescaled skewers, we compute the corresponding fluctuations of the transmitted flux 
$\widetilde{\delta_F}  = ( \widetilde{F} - \langle \widetilde{F} \rangle) / \langle \widetilde{F} \rangle$, where 
$\langle \widetilde{F} \rangle = \exp( - \widetilde{\tau_\mathrm{eff,H}} )$.
Finally, from $\widetilde{\delta_F}$ we compute the mean flux power 
spectrum $\widetilde{P}(k)$ for the rescaled sample.

Figure \ref{fig:rescaled_power_spectrum} shows the fractional difference of the flux power 
spectrum  $\Delta P(k) / P(k) = \widetilde{P}(k)/P(k) - 1 $  
measured between the rescaled skewers and
the original sample for several values 
in the range $\alpha\in[-0.3, 0.3]$.
Because of the non-linear relation between the optical depth $\tau$ and the transmitted flux $F =\exp(-\tau)$, 
rescaling the effective optical depth \taueffH in the skewer sample to higher 
values $\alpha>0$ has the effect of increasing the overall normalization of $P(k)$ on most of
the scales relevant for this work, namely $0.002 \lesssim k \lesssim 0.1 \,\, \mathrm{s\,km^{-1}}$. 
On a similar way, decreasing \taueffH decreases the normalization of $P(k)$ at these scales. For smaller scales 
$k > 0.1 \,\, \mathrm{s\,km^{-1}}$ the effects are redshift dependent and we find that increasing (decreasing) \taueffH tends
to also increase (decrease) $P(k)$ for $z\gtrsim 3.5$, while it has the opposite effect for $z\lesssim 3.5$ as $P(k)$ decreases (increases)
when \taueffH is increased (decreased).

This study shows that the \Lya\ power spectrum itself is sensitive
to the hydrogen effective optical depth, and for this reason
we do not include the observational measurements of \taueffH for our inference of the UVB model presented in this work.

\section{Covariance Matrices of the Transmitted Flux Power Spectrum from the Simulations  } 
\label{sec:covariance_matrices}

In Section \ref{sec:uvb_inference} we present the likelihood function employed for our MCMC analysis (Eq. \ref{eq:mcmc_likelihood}). When
comparing the power spectrum of the \Lya\ transmitted flux from the simulations to the observational measurements
we employ the covariance matrices 
of $P(k)$ reported by the observational works \citep{Chabanier+2019, Irsic+2017a, boera2019a}.
In this section, we quantify 
the effect on the covariance of the simulated $P(k)$ from variations in our model 
parameters.

Figure \ref{fig:covariance_matrix_H} shows the normalized covariance of $P(k)$ at $z=4.6$ for simulations with different 
values for the parameters $\beta_{\mathrm{H}}$ (top panels) and $\Delta z_{\mathrm{H}}$ (bottom panels). 
Decreasing the parameter $\beta_{\mathrm{H}}$ increases the \Lya\ opacity of the IGM,
which increases the normalization of $P(k)$ (see \S\ref{sec:rescale_tau}). The increase of $P(k)$ also increases its covariance on 
roughly all scales. We measure small elementwise differences $<0.1$ in the normalized covariance matrices across simulations that 
vary $\beta_{\mathrm{H}}$, while for simulations with different $\Delta z_{\mathrm{H}}$ the impact is minimal and results in 
only $< 0.03$ elementwise differences. Figure \ref{fig:covariance_matrix_He} presents the covariance matrix of $P(k)$ at $z=3.0$ for 
simulations that vary the parameters $\beta_{\mathrm{He}}$ (top panels) and $\Delta z_{\mathrm{He}}$ (bottom panels). Here we also measure the impact 
to be small with differences $<0.05$.

\begin{figure*}
\includegraphics[width=\textwidth]{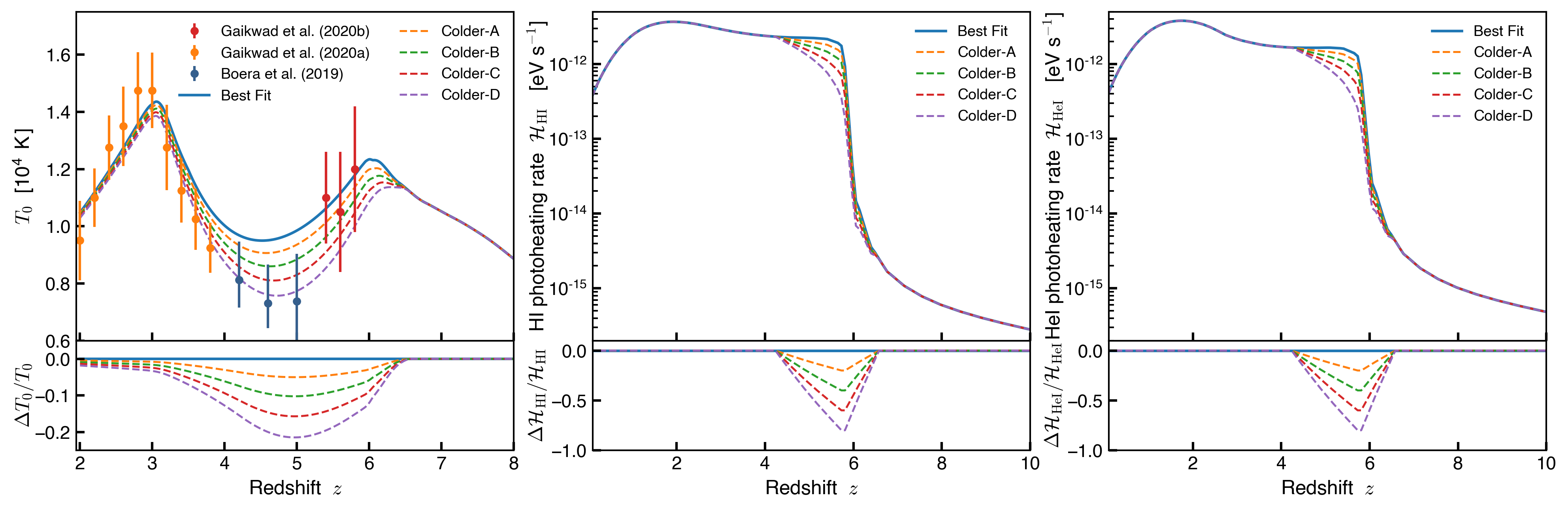}
\caption{Evolution of the IGM temperature $T_0$ (left panel) from models of the UVB where the \HI\ and \HeI\ photoheating rates have been 
reduced in the interval $4.2 \leq z \leq 6.2$ relative to
our best-fit model (center and right panel). The fractional differences of $T_0$ and 
the heating rates $\mathcal{H}_\mathrm{HI}$ and $\mathcal{H}_\mathrm{HeI}$ with respect to the best-fit model are shown in the bottom panels. 
The reduced photoheating rates decrease $T_0$ for $z< 6.2$ but the change is most significant for $3.5 \lesssim z \lesssim 6.0$.  At $z\lesssim 3.5$
the impact on $T_0$ is minimal as heating from \HeII\ reionization dominates.   A reduction of $\sim 80\%$ in the photoheating rates at $z\sim 6.0$ causes 
a decrease in $T_0$ of $\sim 20\%$ at $z \sim  5.0$. For $z \lesssim 3.5$ the reduced photoheating has a minimal impact on $T_0$ of $\lesssim 5\%$.        }
\label{fig:reduced_heating}
\end{figure*}

\section{ Colder Mid-redshift IGM From Reduced Photoheating  }
\label{sec:colder_IGM}

In \S \ref{sec:evolution_temperature } we discuss how our best-fit  model results in a warmer IGM compared to the estimates from \cite{boera2019a} 
during the interval $4.2 \lesssim z \lesssim 5.0$ as the temperature $T_0$ from our model is $\sim 1\sigma$ higher compared to their result.   We 
explore scenarios where the mid-redshift IGM 
is cooled relative to our best-fit model by decreasing the \HI\ and \HeI\ photoheating
rates in the redshift range $4.2 \lesssim z \lesssim 6.2$. The modified photoheating rates are shown in Figure \ref{fig:reduced_heating} (middle and center panels)
along with the fractional differences relative to 
the best-fit model shown in the respective bottom panels.
To compute the history of $T_0$ for the reduced
photoheating models, we integrate the evolution of the temperature of a single-cell at $\rho_\mathrm{gas}=\bar{\rho}$ following the method from \cite{hui1997a} 
(see Section 2 of their work for a detailed description).  The resulting evolution of $T_0$ for the different models is presented in the left panel of Figure \ref{fig:reduced_heating}.  
We show that reducing the \HI\ and \HeI\ photoheating rates by $\sim 80\%$ at $z\sim 6$ results in a colder IGM where $T_0$ is reduced by $\sim 20\%$ at $z\sim 5$ 
such that $T_0 \sim 8\times 10^3\, \mathrm{K}$ for $4.2 \lesssim z \lesssim 5.0$ agrees well with the estimate from \cite{boera2019a}.
However, we find that for such a scenario
$T_0$ at $z \sim 5.4$ is lower than the inference from \cite{gaikwad2020a}. This conflict exhibits some degree of tension between the estimates  at $z\sim 5.0$ and 
$z \sim 5.4$ from \cite{boera2019a} and \cite{gaikwad2020a} respectively.

The photoheating $\mathcal{H}$ and the photoionization $\Gamma$ rate from the UVB are given by the intensity  of the background radiation $J(\nu, z)$ as

\begin{equation}
\Gamma(z)=\int_{\nu_{0}}^{\infty} \frac{4 \pi J(\nu, z)}{h \nu} \sigma(\nu) d \nu, \;\;\;\; \;\;\;\;
\mathcal{H}(z) =\int_{\nu_0}^{\infty} \frac{4 \pi J(\nu, z)}{h \nu}\left(h \nu-h \nu_{0}\right) \sigma(\nu) d \nu
\label{eq:photoionization_photoheating}
\end{equation}

\noindent
where $\nu_0$ and $\sigma(\nu) $ are the threshold  frequency and  photoionization cross-section, respectively. 
Consider power-law models for the cross-section and the intensity
of the radiation at wavelengths $\lambda > 912 \, \textup{\AA}$, which
can be written 
as $\sigma(\nu) = \sigma_0 (\nu/\nu_0)^\phi$ and $J(\nu) =  (\nu/\nu_0)^{\alpha}$, with 
indices $\phi<0$ and $\alpha <0$. 
Physically, reducing the photoheating rate relative
to the photoionization rate can be achieved by changing the spectral index  of the ionizing radiation
$\alpha$.  By solving the integrals in Eqs \ref{eq:photoionization_photoheating} assuming
these power-law models
and evaluating the fractional change in the photoionization $\Delta \Gamma / \Gamma$ 
and photoheating $\Delta \mathcal{H} / \mathcal{H}$ for a change in the spectral index $\Delta \alpha$, we find the following relation is satisfied

\begin{equation}
\Delta \alpha=(1+\alpha+\phi) \frac{\frac{\Delta \Gamma}{\Gamma}-\frac{\Delta \mathcal{H}}{\mathcal{H}}}{1+\frac{\Delta \mathcal{H}}{\mathcal{H}}}
\label{eq:Gamma_H_alpha_relation}
\end{equation}

\noindent Equation \ref{eq:Gamma_H_alpha_relation} relates the change of the spectral index of the radiation necessary to produce some variation of the photoionization and photoheating 
from a given UVB model. By applying Eq. \ref{eq:Gamma_H_alpha_relation},
we can modify the photoheating relative to
the photoionization of a UVB model within a 
physically-plausible range for the index $\alpha$.
In future work, we will explore which variations in the IGM temperature $T_0$ from changes of the photoheating rate match the
observed hydrogen effective optical depth at $z>5$ while using physically-plausible source
populations.

\begin{figure*}
\includegraphics[width=\textwidth]{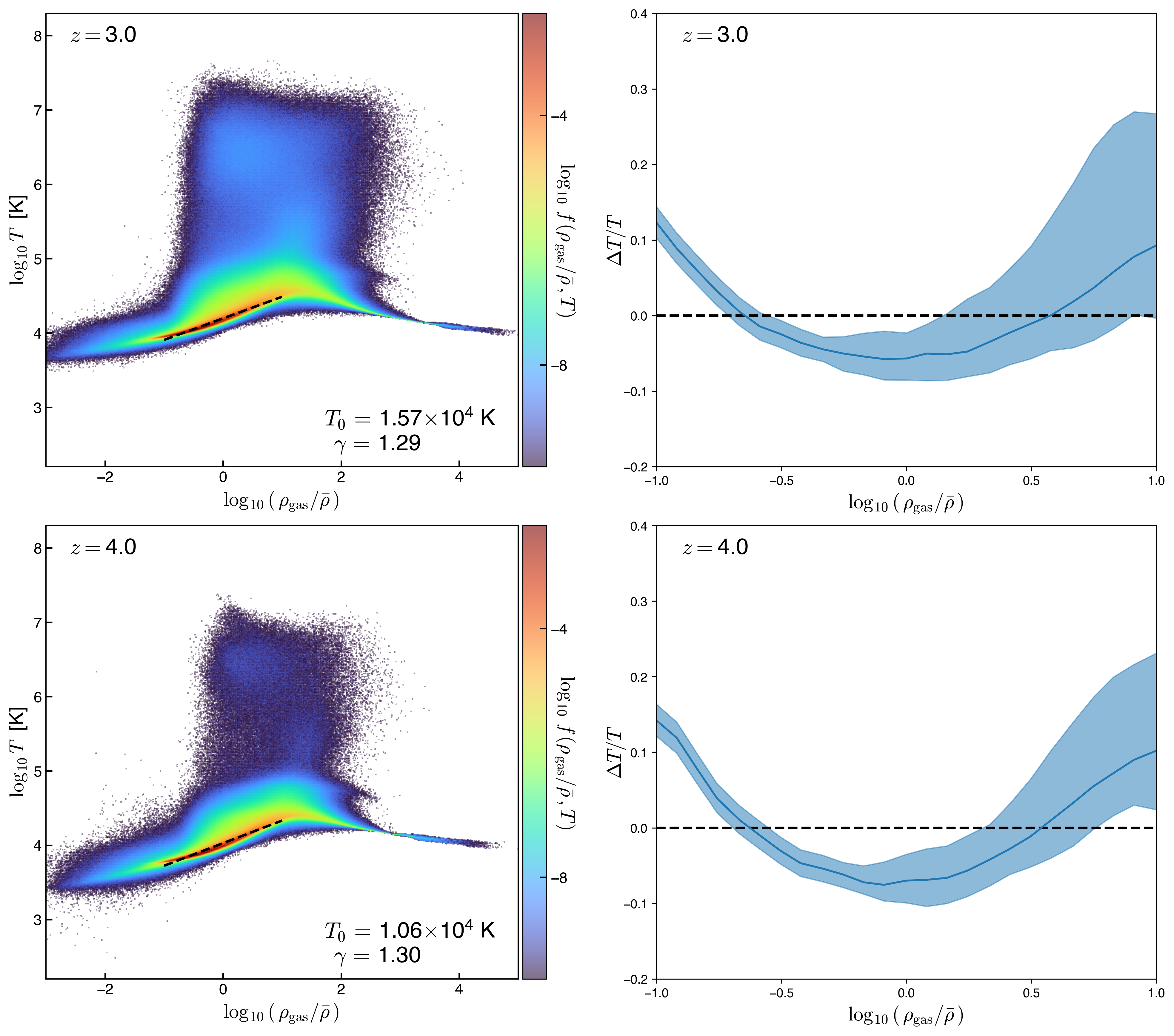}
\caption{Density-temperature distribution of the IGM gas (left column) from one of our simulations at
redshift $z=3$ (top) and $z=4$ (bottom). 
A power law fit of the form $T=T_0 \left( \rho_\mathrm{gas} / \bar{\rho} \right)^{\gamma-1}$ over the range $-1 \leq  \log_{10}(\rho_\mathrm{gas} / \bar{\rho} ) \leq 1$ 
is shown (black dashed lines). The right columns shows 
deviations of the density-temperature distribution
with respect to the power-law fit over the fitted range. 
The blue region corresponds to the 68\% highest probability interval for the temperature as function of the overdensity $\rho_\mathrm{gas} / \bar{\rho}$. 
The differences between the distribution of the gas
relative to
the power-law fit can be as large as $\sim 15\%$.}
\label{fig:phase_diagrams}
\end{figure*}

\section{ Accuracy of the Power-Law fit to the Density-Temperature Distribution of the Gas in Our Simulations. }
\label{sec:phase_diagram}

A common method 
to infer the thermal state of the IGM from observations of the \Lya\ forest
involves marginalizing over the thermal properties $T_0$
and $\gamma$ in the 
approximate power-law density-temperature relation $T=T_0 \left( \rho_\mathrm{gas} / \bar{\rho} \right)^{\gamma-1}$ \citep{ bolton2014a, Nasir+2016, hiss2018a,  boera2019a,   walther2019a, gaikwad2020b}.
The density of the 
IGM gas that contributes to the majority of the \Lya\ forest signal lies
in the range  $-1 \leq  \log_{10}(\rho_\mathrm{gas} / \bar{\rho} ) \leq 1$. From our simulations
we find that a single power law fails to reproduce the density-temperature distribution of the gas over this density interval. The left panels of 
Figure \ref{fig:phase_diagrams} show 
the density-temperature distribution of the gas in one of our simulations and the corresponding power-law fit to the distribution over the density 
range $-1 \leq  \log_{10}(\rho_\mathrm{gas} / \bar{\rho} ) \leq 1$ at redshift $z=3$ (top) and $z=4$ (bottom). The deviations of the gas 
temperature in the simulation relative to
the power-law fits are
presented in the right panels, showing that the 
fractional differences $\Delta T /T$ from the density-temperature distribution in the simulation with 
respect to the power-law fit can be as large as $\sim 15\%$.

\bibliography{references}

\end{document}